\documentclass[pdflatex,sn-mathphys-num]{sn-jnl}

\usepackage{graphicx}%
\usepackage{multirow}%
\usepackage{amsmath,amssymb,amsfonts}%
\usepackage{amsthm}%
\usepackage{mathrsfs}%
\usepackage[title]{appendix}%
\usepackage{xcolor}%
\usepackage{textcomp}%
\usepackage{manyfoot}%
\usepackage{booktabs}%
\usepackage{algorithm}%
\usepackage{algorithmicx}%
\usepackage{algpseudocode}%
\usepackage{float}%
\usepackage{listings}%
\usepackage{enumitem}
\usepackage{subcaption}
\usepackage{bm}

\begin{document}

\title[Anisotropy in Pantheon+ supernovae]{Anisotropy in Pantheon+ supernovae}


\author*[1]{\fnm{Animesh} \sur{Sah}}\email{animesh.sah@tifr.res.in}

\author[1]{\fnm{Mohamed} \sur{Rameez}}\email{mohamed.rameez@tifr.res.in}

\author[2]{\fnm{Subir} \sur{Sarkar}}\email{subir.sarkar@physics.ox.ac.uk}

\author[3,4]{\fnm{Christos\sur{Tsagas}}\email{tsagas@astro.auth.gr}}

\affil*[1]{\orgdiv{Department of High Energy Physics}, \orgname{Tata Institute of Fundamental Research}, \orgaddress{\street{Homi Bhabha Road}, \city{Mumbai}, \postcode{400005},\country{India}}}

\affil[2]{\orgdiv{Rudolf Peierls Centre for Theoretical Physics}, \orgname{University of Oxford}, \orgaddress{\street{Parks Road}, \city{Oxford}, \postcode{OX1 3PU}, \country{United Kingdom}}}

\affil[3]{\orgdiv{Section of Astrophysics, Astronomy \& Mechanics}, \orgname{Aristotle University of Thessaloniki}, \orgaddress{\street{}, \city{Thessaloniki}, \postcode{54124}, \country{Greece}}}

\affil[4]{\orgdiv{Clare Hall}, \orgname{University of Cambridge}, \orgaddress{\street{Herschel Road}, \city{Cambridge} \postcode{CB3 9AL}, \country{United Kingdom}}}


\abstract{We employ Maximum Likelihood Estimators to examine the Pantheon+ catalogue of Type Ia supernovae for large scale anisotropies in the expansion rate of the Universe. The analyses are carried out in the heliocentric frame, the CMB frame, as well as the Local Group frame. In all frames, the Hubble expansion rate in the redshift range $0.023 < z < 0.15$ is found to have a statistically significant dipolar variation exceeding 1.5 km\,s$^{-1}$\,Mpc$^{-1}$, i.e. bigger than the claimed 1\% uncertainty in the SH0ES measurement of the Hubble parameter $H_0$. The deceleration parameter too has a redshift-dependent dipolar modulation at $>5\sigma$ significance, consistent with previous findings using the SDSSII/SNLS3 Joint Lightcurve Analysis catalogue. The inferred cosmic acceleration cannot therefore be due to a Cosmological Constant, but is likely a general relativistic effect due to the anomalous bulk flow in our local Universe.}


\newcommand\blfootnote[1]{
    \begingroup
    \renewcommand\thefootnote{}\footnote{#1}
    \addtocounter{footnote}{-1}
    \endgroup
}

\maketitle

\section{Introduction} 
\label{sec:intro}

The standard `Lambda Cold Dark Matter' ($\Lambda$CDM) cosmological model is based on the Friedmann-Lema\^itre-Robertson-Walker (FLRW) metric which describes maximally symmetric space-time. The Friedmann-Lema\^itre equations hold in the `cosmic rest frame' (CRF) in which all galaxies recede from each other in the Hubble expansion and the cosmic microwave background (CMB) looks isotropic. In the real Universe, the CMB has however a pronounced dipole anisotropy; this is interpreted as due to our `peculiar' motion (because of local inhomogeneities) with respect to the CRF \cite{Stewart:1967ve}. Accordingly cosmological observables, e.g. redshifts and apparent magnitudes of Type Ia supernovae (SNe~Ia), as measured in our heliocentric frame, are boosted to the CRF to be analysed in the FLRW framework. Data on SNe~Ia \cite{Brout:2022vxf}, small-angle CMB anisotropies \cite{Planck:2018nkj}, galaxy clustering \& weak lensing \cite{DES:2017myr}, baryon acoustic oscillations \cite{DESI:2024mwx} etc, are all analysed \emph{assuming} FLRW and are concordant with the $\Lambda$CDM model.

The FLRW assumption is however now challenged because the dipole anisotropy measured in the sky distribution of cosmologically distant  sources does \emph{not} match that expected from the kinematic interpretation of the CMB dipole~\cite{1984MNRAS.206..377E}. This crucial consistency test, performed with multiple independent flux limited surveys in radio and infrared, from both ground and space based observatories, rejects the FLRW assumption at over $5\sigma$~\cite{Secrest:2022uvx, Dam:2022wwh,Wagenveld:2023kvi}. Moreover, longstanding indications of a large-scale coherent component to peculiar motions in the local Universe, first traced out to $\sim200 h^{-1}$~Mpc using SNe~Ia as independent distance indicators \cite{Colin:2010ds,Feindt:2013pma}, has recently been confirmed by the CosmicFlows-4 survey using Tully-Fisher and Fundamental Plane distances; this bulk flow deviates from the $\Lambda$CDM expectation by $ 4-5\sigma$ \cite{Watkins:2023rll,Whitford:2023oww}. These anomalies have however received less attention than the discrepancy, within the FLRW framework, between the Hubble parameter $H_0$ inferred from CMB anisotropies~\cite{Planck:2018nkj} and determined from the SNe~Ia Hubble diagram in the local Universe~\cite{Riess:2021jrx}.

Additionally, when the SDSS-II/SNLS-3 Joint Lightcurve Analysis (JLA) catalogue of SNe~Ia~\cite{SDSS:2014iwm} was analysed in the heliocentric frame (after removing demonstrably incorrect peculiar velocity corrections), the inferred acceleration of the expansion was found to be \emph{anisotropic}, with the dipole component of the deceleration parameter $q_0$ dominating over its monopole component 
out to redshift $z \sim 0.1$~\cite{Colin:2019opb}. Acceleration due to $\Lambda$ must be isotropic, so this rejects the $\Lambda$CDM model, at $3.9\sigma$. 
Such a dipole is a characteristic signature of relative motion, so the observed \emph{anisotropic} acceleration is likely a general relativistic effect due to our observations being carried out from within the local bulk flow which is contracting, especially since the dipole in $q_0$ is seen to decay with redshift as is expected for a decaying flow~\cite{Tsagas:2009nh,Tsagas:2011wq}. \footnote{Analysis of the peculiar velocity field reconstructed from the 2M++ survey \cite{Carrick:2015xza} indicates that the local bulk flow is contracting on average \cite{Pasten:2023rpc}. However  CosmicFlows-4 data shows the bulk velocity still rising out to $\sim200h^{-1}$~Mpc \cite{Watkins:2023rll}; this may be due to cosmic variance, since the velocity profile around an individual observer need not decrease monotonically, even though the ensemble average does so \cite{Mohayaee:2020wxf}.}

These developments warrant a reexamination of the latest public compilation of SNe~Ia data, the Pantheon+ catalogue~\cite{Scolnic:2021amr} described in \S~\ref{sec:dataset}. In \S~\ref{sec:floats}, we outline our analysis which, as in previous work \cite{Colin:2019opb}, is based on a cosmographic expansion of the luminosity distance in redshift, using an unbiased Maximum Likelihood Estimator \cite{Nielsen:2015pga}.
We scrutinise the data for any anisotropy in the Hubble parameter $H_0$ or the deceleration parameter $q_0$ and detect a significant dipole asymmetry in both. In \S~\ref{sTCs} we discuss how this might arise due to our being `tilted observers' embedded in a fast and deep bulk flow. In \S~\ref{discuss} we compare our results with related work, and present our conclusions in \S~\ref{concl}.

\section{The Dataset - Pantheon+} \label{sec:dataset} 

\begin{figure}[htb]
    \centering
    \includegraphics[width=1.0\columnwidth]{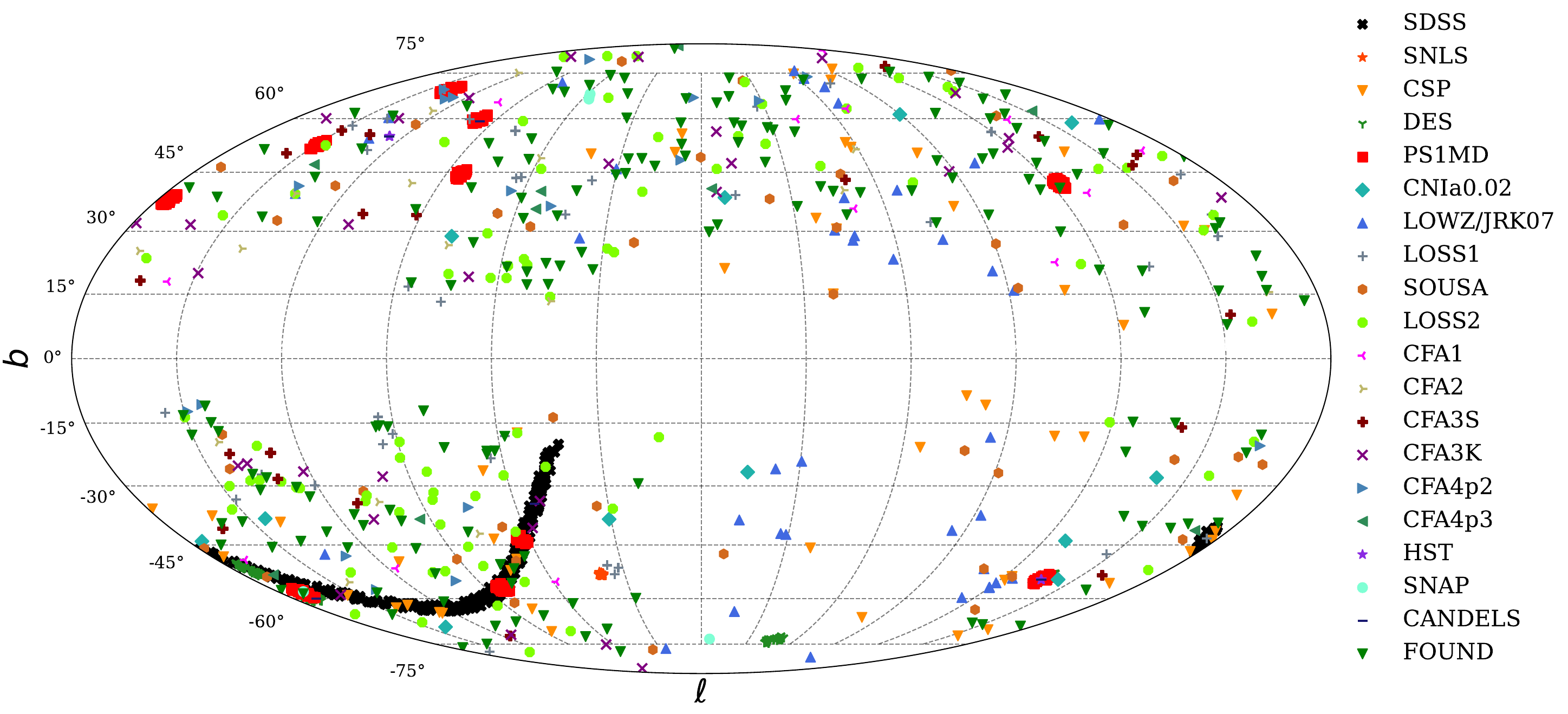}
    \caption{Mollweide sky map of all 1533 unique Type Ia supernovae in the Pantheon+ catalogue, plotted in Galactic co-ordinates. The colours refer to the individual sub-catalogues, listed in Ref.~\cite{Scolnic:2021amr}.}
    \label{fig:skyview}
\end{figure}

\begin{figure}[ht!]
    \centering
\includegraphics[width=0.5\columnwidth]{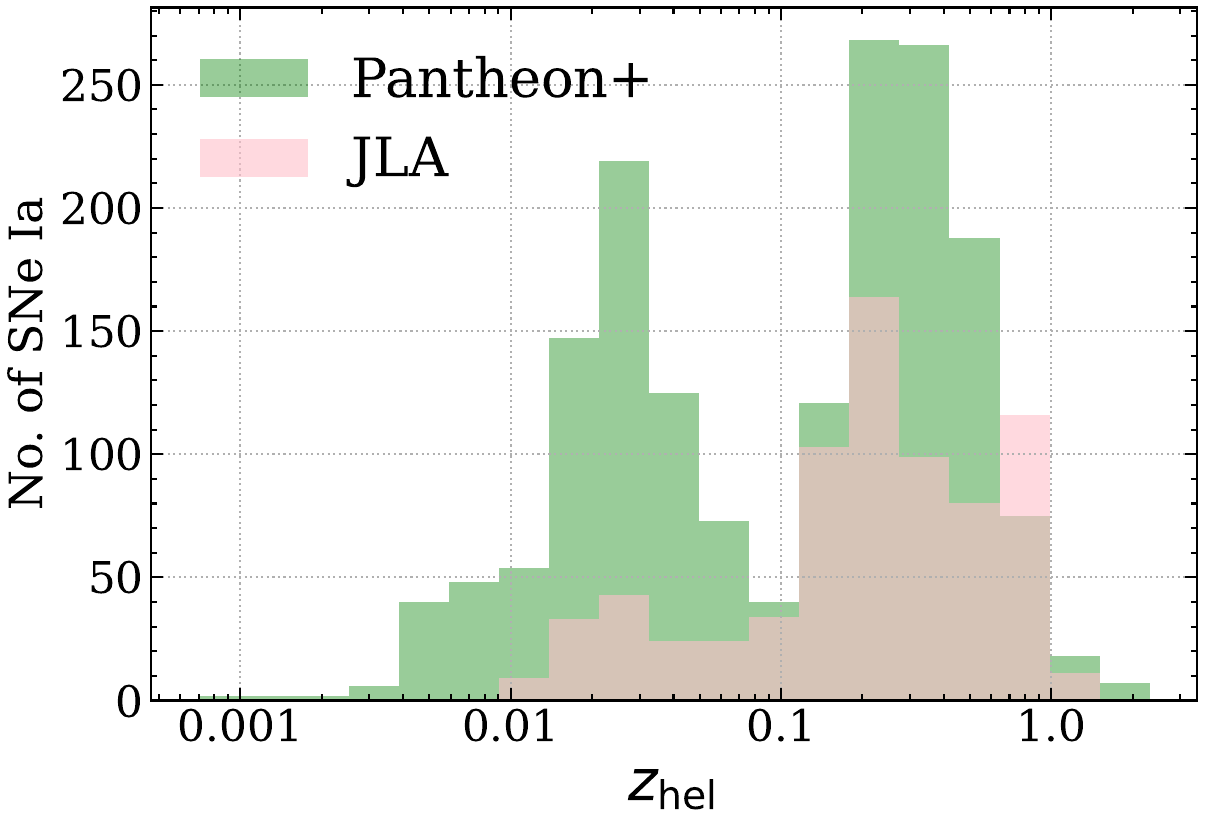}
    \caption{Distribution of heliocentric redshifts in  the Pantheon+ SNe~Ia catalogue~\cite{Scolnic:2021amr}, compared to the SDSS-II/SNLS3 Joint Lightcurve Analysis (JLA) catalogue~\cite{SDSS:2014iwm}.}
    \label{fig:zhist}
\end{figure}

The Pantheon+ compilation\cite{Scolnic:2021amr} (see Figs.~\ref{fig:skyview} \& \ref{fig:zhist}) consists of data from 1701 observations of spectroscopically confirmed SNe~Ia amalgamated from many different surveys, of which we find 1533 are unique events (i.e. not multiple observations of the same supernova). Of the 740 SNe~Ia which made up the JLA sample~\cite{SDSS:2014iwm}, 584 were included in Pantheon+ after quality cuts. Of the additional 949 SNe~Ia, 446 have been added at redshift $z < 0.1$ as seen in Fig.~\ref{fig:zhist}; of these, 66 are below the lowest redshift ($z=0.00937$) in the JLA catalogue. 

The Pantheon+ catalogue, like JLA, is disseminated with the data already corrected for peculiar velocities~\cite{Carr:2021lcj} --- both our velocity (in the heliocentric frame) of $369.8$~km~s$^{-1}$ w.r.t. the CRF \cite{Planck:2018nkj}, as well as that of the host galaxy of each supernova w.r.t. the CRF. The model~\cite{Carrick:2015xza} employed for making these corrections infers the peculiar velocity field from the 2M++ density field using linear Newtonian perturbation theory; according to this model, a bulk flow of $159 \pm 23$ km s$^{-1}$ continues outside the survey volume ($r>r_\text{max} \sim 200h^{-1}\text{Mpc}$) at $>5 \sigma$ significance. Unlike in JLA, the peculiar velocity corrections in Pantheon+ do not terminate at $r_\text{max}$ which would introduce an unphysical discontinuity as noted earlier \cite{Colin:2019opb}; rather, to ensure a smooth transition the bulk flow is modelled as a decaying function according to the $\Lambda$CDM expectation~\cite{Carr:2021lcj}, thereby extending the peculiar velocity corrections to \emph{all} SNe~Ia in the catalogue.\footnote{Peculiar velocities $v$ are defined as the residual velocities of objects at distance $d$, after the isotropic Hubble expansion is subtracted out, i.e. $v = cz - H_0 d$. Note that this decomposition is a  Newtonian concept.} Note that although the peculiar velocity corrections are small in amplitude relative to the Hubble flow beyond $r_\text{max}$, they are directionally coherent.

Pantheon+ employs different terminology from JLA; while the redshifts corrected for \emph{both} observer and SNe~Ia motion were called $z_\text{CMB}$ in JLA, they are called $z_\text{HD}$ in Pantheon+, referring to the redshifts used to construct the `Hubble Diagram' (HD); note that this is just the usual redshift $z$ in the CRF of the assumed FLRW model i.e. $z_\text{HD} \equiv z$. Now $z_\text{CMB}$ refers to redshifts which are corrected \emph{only} for the observer motion w.r.t. the CRF. Unlike JLA, the covariance matrix for each source of correlated systematic uncertainties, e.g. calibration, dust, extinction and peculiar velocities, are no longer provided separately, however these can be obtained  following Ref.~\cite{Lane:2023ndt}. We note that the diagonal covariance terms in the Pantheon+ covariance matrices are much larger than in JLA (see Fig.~\ref{fig:histcomp}). 

\section{Analyses} \label{sec:floats} 

The Pantheon+ analysis~\cite{Brout:2022vxf} follows previous work (e.g. Refs.~\cite{SupernovaCosmologyProject:1998vns,SNLS:2005qlf,SDSS:2014iwm}) in using the `constrained $\chi^2$' statistic in which  error bars are \emph{adjusted} until a good fit with $\chi^2$/d.o.f=1 is obtained. This is however quite unsuitable for goodness-of-fit testing or model selection~\cite{March:2011xa}. We use therefore the Maximum Likelihood Estimator (MLE) constructed to enable a statistically principled treatment of the intrinsic scatter in SNe~Ia data~\cite{Nielsen:2015pga}. Our approach is frequentist, but completely equivalent to employing a  Bayesian Hierarchical model~\cite{March:2011xa,Shariff:2015yoa,Dam:2017xqs,Lane:2023ndt}.

We employ the cosmographic Taylor expansion for the luminosity distance $d_{L}$ (related to the distance modulus as $\mu \equiv 25 + 5\text{log}_{10}(d_{L}/\text{Mpc})$) to 3rd-order in redshift, in terms of the Hubble velocity ($H_0 \equiv ({\dot{a}/a})|_{z=0}$), deceleration ($q_0 \equiv ({a\ddot{a}/\dot{a}^2})|_{z=0}$) and jerk ($j_0 \equiv ({a^2\dddot{a}/\dot{a}^3})|_{z=0}$) parameters~\cite{Weinberg:1972kfs,Visser:2003vq}, where the overdot signifies the time derivative $\text{d}/\text{d}t$:
\begin{equation}
d_{L}(z)= \frac{cz}{H_0}\left[{1+\frac{1}{2}(1-q_0)z - \frac{1}{6}\left(1-q_0 - 3q_0^2 +j_0+ \frac{kc^2}{H_0^2 a_0^2}\right)z^2}\right],
\end{equation}
which is modified to account for our local peculiar velocity thus~\cite{Rubin:2019ywt}:
\begin{equation}
d_{L}(z, z_\text{hel}) = \frac{cz}{H_0}\left[{1+\frac{1}{2}(1-q_0)z - \frac{1}{6}\left(1 - q_0 - 3q_0^2 +j_0 + \frac{kc^2}{H_0^2 a_0^2}\right)z^2}\right] \times \frac{1+z_\text{hel}}{1+z},
\label{lum_dist}
\end{equation}
where $z_\text{hel}$ refers to the measured redshift  in the heliocentric frame. This ensures that $d_{L} = (1+z_\text{hel})\times$(comoving distance)~\cite{Rubin:2019ywt} and we use this for the present analysis.
For FLRW cosmologies, $q_0=\Omega_\text{m}/2-\Omega_{\Lambda}$, hence the fiducial flat $\Lambda$CDM model with $\Omega_\text{m}=0.315$, $\Omega_{\Lambda}=0.685$ \cite{ParticleDataGroup:2024cfk} has $q_0=-0.528$ and $j_0=1$~\cite{Rapetti:2006fv}.
To ensure good convergence (within 2\%) of the Taylor expansion, we consider only supernovae at $z_\text{hel} < 0.8$, thereby excluding 31 SNe~Ia and leaving 1670 for our analysis. 

The covariance matrix in Pantheon+ differs from JLA, in providing only an `$N\times N$' covariance matrix, where $N$ is the number of SNe~Ia. By contrast, JLA provided the full `$3N\times 3N$' covariance matrix which included for each supernova the covariance between the observed peak magnitude $m_B$ (in the rest frame `$B$'-band) and the `stretch' ($x_1$) and `colour' ($c$) standardisation made in the Spectral Adaptive Lightcurve Template 2 (SALT2) fitter \cite{SNLS:2007cqk}. The SALT2 parameters yield the  distance modulus via the Phillips-Tripp formula \cite{Phillips:1993ng,Tripp:1997wt}:
\begin{equation}
\mu = m_{B\text{corr}} - M - \delta_{\text{bias}} +\delta_{\text{host}}, \quad m_{B\text{corr}} = m_B + \alpha x_1 - \beta c,
\label{Phillips-Tripp}
\end{equation}
with $M$ being the absolute magnitude, $\delta_{\text{bias}}$ the `selection bias correction' and $\delta_{\text{host}}$ the `host-galaxy correction'.  The JLA covariance had the form~\cite{SDSS:2014iwm}:
\begin{equation}
C^\text{JLA}_{i} = 
\begin{pmatrix}
  \sigma^2_{m_{B_i}} & \sigma_{m_{B},x_{1_i}} &\sigma_{m_{B}, c_i} \\
  \sigma_{m_{B},x_{1_i}} & \sigma^2_{x_{1_i}} & \sigma_{x_1,c_i} \\ 
  \sigma_{m_{B},c_i} & \sigma_{x_1,c_i} & \sigma^2_{c_i}
\end{pmatrix},
\end{equation}
whereas the Pantheon+ covariance matrix is: $C^\text{Pantheon+} (i, j) = C_\text{stat}^\text{Pantheon+}  (i, j) + C_\text{syst}^\text{Pantheon+} (i, j)$, with the statistical part being \cite{Brout:2022vxf}:
\begin{equation} 
C^\text{Pantheon+}_{\text{stat}} (i, j) = 
\left\{
\begin{array}{ll}
\sigma^2_{\mu} & \quad i = j, \\
\sigma^2_{\text{floor}} + \sigma^2_{\text{lens}} + \sigma^2_{z} +\sigma^2_{v_\text{pec}}& \quad i \neq j \text{ and SN}_{i} = \text{SN}_{j}. \\
\end{array}
\right.
\label{covar}
\end{equation}
Here the uncertainty in the distance modulus is:
\begin{equation} 
\sigma^2_{\mu, i}  = f(z_i,c_i,M_{*,i})\sigma^2_{\text{meas,i}}+
\sigma^2_{\text{floor}}(z_i,c_i,M_{*,i}) + \sigma^2_{\text{lens},i} + \sigma^2_{z,i} + \sigma_{v_\text{pec,i}}, \\
\label{eq_mu}
\end{equation} 
where $\sigma_{\text{meas}}$ denotes the measurement uncertainty of SALT2 light curve fits (scaled by a factor $f_{z_i,c_i,M_{*,i}}$ to account for selection effects) and $\sigma_\text{floor}$, $\sigma_\text{lens}$, $\sigma_z$ and $\sigma_{v_\text{pec}}$ represent, respectively, the floor in standardisability (due to intrinsic unmodeled variations in SNe~Ia),  gravitational lensing uncertainty, redshift measurement uncertainty, and  peculiar velocity uncertainty.

 We remove the selection bias corrections that were applied to the SNe~Ia magnitudes as these were calculated assuming the $\Lambda$CDM model. As stated in Ref.~\cite{Brout:2022vxf}, the bias corrections $\delta_\text{bias}$ added to $m_{B\text{corr}}$ in Eq.~(\ref{Phillips-Tripp}) in Pantheon+ are of \emph{opposite} sign to the corrections made in JLA so in our case these are added.\footnote{If the bias corrections are not removed, the statistical significance of the dipole in $q_0$ remains unchanged (within $1\sigma$), however its monopole component becomes more negative in all frames.}

Pantheon+ provides the `Phillips-Tripp corrected' magnitude $m_{B\text{corr}}$ which can be used directly with the provided `$N\times N$' covariance matrix. Note however that this light curve standardisation has been done allowing $x_1$ and $c$ to be both sample and redshift-dependent, as advocated in Refs.~\cite{Rubin:2019ywt,Rubin:2016iqe}, even though this undermines the case for assuming that $M$ is \emph{not} dependent on  redshift, i.e. that SNe~Ia are indeed standard(isable) candles~\cite{Mohayaee:2021jzi}.

To illustrate the effects of such \emph{ad hoc} corrections and enable comparison with results obtained using different methodologies, we carry out two types of analyses:\\

\begin{trivlist}
 
\item \textbf{C1}: We use the `$N \times N$' ($N$ being the number of SNe~Ia) statistical+systematic covariance matrix and the (sample and redshift-dependent) `Phillips-Tripp corrected' magnitude $m_{B\text{corr}}$ (\ref{Phillips-Tripp}) provided with the Pantheon+ dataset \cite{Brout:2022vxf}. However, rather than the statistically unprincipled `constrained $\chi^2$' method, we use the MLE~\cite{Nielsen:2015pga} with the likelihood:
\begin{equation}
\mathcal{L}[\theta]=\int p[\hat{m}_{B\text{corr}}|M] \times p[M|\theta] \text{d}M ,
\end{equation}
where the hat refers to the observed quantity, and $p$ to the underlying probability distribution of the true data. Here $p[M|\theta]$ is taken to be a Gaussian (following Ref.~\cite{Nielsen:2015pga}) with $M_0$ as the central value and $\sigma_{M_0}$  its variance: 
$p(M|\theta)=(2\pi\sigma^2_{M_{0}})^{-1/2} \exp(\{-[(M-M_{0})/\sigma_{M_{0}}]^2/2\})$.

The likelihood is then simply obtained by analytic integration \cite{Nielsen:2015pga}:
\begin{equation}
    \mathcal{L} = |2 \pi (\Sigma_d +A^T \Sigma_l )|^{-1/2} \times 
    \exp{[-(\hat{Z}-Y_0 )( \Sigma_d +A^T \Sigma_l )^{-1} (\hat{Z}-Y_0)^T/2]},
\end{equation}
where the vector $\hat{Z} = \{\hat{m}_{B\text{corr}} - \mu_1,...\}$, $\mu_i$ being the distance modulus estimated from theory. Here $Y_0 = \{M_0,...\}$  is the vector of the mean of the Gaussian model of $M$, 
$\Sigma_d$ is the covariance matrix (which includes both statistical and systematic terms) and $\Sigma_l = \text{diag}(\sigma^2_{M0},...)$ is the matrix consisting of the variance of the Gaussian model.

Analysis C1 thus involves evaluating the cosmological parameters ($q_0$, $j_0-\Omega_k$), as well as $M_0$ and $\sigma_{M_0}$.\\

\item \textbf{C2}: In order to propagate the uncertainties from the SNe~Ia light-curve correction process to cosmological parameter inference more faithfully, we perform a second type of analysis, employing the apparent magnitude $m_B$ (which is \emph{not} `Phillips-Tripp corrected'). We use the covariance matrix from Ref.~\cite{Lane:2023ndt} which is a `$3N \times 3N$' covariance matrix built using the SALT2 variances from the Pantheon+ dataset. Since some elements make the covariance non-`positive semi-definite', 17 SNe~Ia (of which 15 are unique) are removed,\footnote{In fact there is one more, but it is in the already excluded set having $z>0.8$.} leaving 1653 for our analysis. This covariance has the structure: 
\begin{equation}
C_{i} = 
\begin{pmatrix}
  \sigma^2_{m_{B_i}} & \sigma_{m_{B},x_{1_{i}}} &\sigma_{m_{B}, c_i} \\
  \sigma_{m_{B},x_{1_i}} & \sigma^2_{x_{1_i}} & \sigma_{x_1,c_i} \\ 
  \sigma_{m_{B}, c_i} & \sigma_{x_1,c_i} & \sigma^2_{c_i}
\end{pmatrix},
\label{eq:3n3nmatrix}
\end{equation} 
and is constructed by using the SALT2 uncertainties from the Pantheon+ dataset. Furthermore the lensing variance $\sigma^2_\text{lens} (= 0.055 z_{\text{CMB}})$ \cite{Jonsson:2010wx} and the redshift variance $\sigma_{z}$ are added to the first element of each diagonal block. Finally, the systematic and covariance due to duplicates are added separately. Thus the peculiar velocity and other model-\emph{dependent} variances are not included, unlike the covariance matrix provided with the Pantheon+ dataset \cite{Brout:2022vxf}.\\

\end{trivlist}

The likelihood is then \cite{Nielsen:2015pga}:
\begin{equation} 
\label{eq4}
\mathcal{L[\theta]} = p[(\hat m_{B}, \hat x_1 ,\hat c  )|\theta]  
                    = \int{p[(\hat m_{B}, \hat x_1, \hat c)|(M, x_1, c)] \times p[(M,x_1 , c)|\theta]\,\text{d}M \text{d}x_1 \text{d}c}, 
\end{equation}
where the hats above refer to the observed quantities and $p$ to the underlying probability distributions of the true data. Here $p[x_1|\theta]$ and $p[c|\theta]$ are taken to be Gaussian with mean $x_{1,0}$ and $c_0$, and variances $\sigma_{x_{1,0}}$ and $\sigma_{c_{0}}$, respectively. 
For the analysis C2, we employ sample and redshift-\emph{independent} Gaussian parametrisations for both $x_1$ and $c$~\cite{Colin:2019opb,Nielsen:2015pga}: $p(x_1|\theta)=(2\pi\sigma^2_{x_{1,0}})^{-1/2} \exp(\{-[(x_1-x_{1,0})/\sigma_{x_{1,0}}]^2/2\})$, 
 $p(c|\theta)=(2\pi\sigma^2_{c_{0}})^{-1/2} \exp(\{-[(c-c_{0})/\sigma_{c_{0}}]^2/2\})$, and  $p(M|\theta)=(2\pi\sigma^2_{M_{0}})^{-1/2} \exp(\{-[(M-M_{0})/\sigma_{M_{0}}]^2/2\})
$.
The likelihood is again  obtained by analytic integration \cite{Nielsen:2015pga}:
\begin{equation}
    \mathcal{L} = |2 \pi (\Sigma_d +A^T \Sigma_l A)|^{-1/2} \times 
    \exp{[-(\hat{Z}-Y_0 A)( \Sigma_d +A^T \Sigma_l A)^{-1} (\hat{Z}-Y_0A)^T/2]},
\end{equation}
where the vector $\hat{Z} = \{\hat{m}_{B1} - \mu_1,\hat{x}_1,\hat{c},...\}$, $Y_0 = \{M_0,x_{1,0},c_0,...\}$  is the vector of the mean of the Gaussian models and \[
A =
\begin{pmatrix}
1 & 0 & 0 &  \\
-\alpha & 1 & 0 & 0 \\
\beta & 0 & 1 &  \\
 &0 & & \ddots
\end{pmatrix}.
\]
Here $\Sigma_d$ is the covariance matrix (which includes both statistical and systematic terms) and $\Sigma_l = \text{diag}(\sigma^2_{M0},\sigma^2_{x_{1,0}},
\sigma^2_{c_0},...)$ is the matrix consisting the variance of the Gaussian models.
Analysis C2 thus involves evaluating the cosmological parameters ($q_0$,~$j_0-\Omega_k$), and simultaneously all the Phillips-Tripp light curve standardisation parameters ($\alpha$, $\beta$, $x_{1,0}$, $\sigma_{x_{1,0}}$, $c_0$, $\sigma_{c_0}$) as well as $M_0$ and $\sigma_{M_0}$.

In addition to $z_\text{hel}$ and $z_\text{CMB}$, the redshifts in the heliocentric and CMB frames, we also consider $z_\text{LG}$, the redshift boosted to the Local Group (LG) frame. This is obtained using \cite{Davis:2010jq,Ellis:1987zz}: 
\begin{equation} 
\label{eqzmod}  
(1+z_\text{LG}) = (1+z_\text{hel}) \times (1+z_{\text{LG-hel}})
\end{equation}
where $z_{\text{LG-hel}}= \sqrt{\frac{1-\vec{v}_{\text{LG-Sun}}\cdot \hat{n}/c}{1+\vec{v}_{\text{LG-Sun}}\cdot \hat{n}/c}} -1$ and $\vec{v}_{\text{LG-Sun}}$ is the velocity of the Local Group frame relative to the heliocentric frame. 
We adopt the values given in Table 3 of Ref.~\cite{Planck:2018nkj}. The Sun's motion around the Galaxy is roughly in the \emph{opposite} direction to the CMB dipole hotspot, so while the heliocentric frame moves wrt the CMB at $369.82 \pm 0.11$~km\,s$^{-1}$ towards $l~\text{[deg]} = 264.021 \pm 0.011, b~\text{[deg]} = 48.243 \pm 0.005$, the LG moves wrt the CMB at $620 \pm 15$~km\,s$^{-1}$ towards $l~\text{[deg]} = 271.9 \pm 2.0, b~\text{[deg]} = 29.6 \pm 1.4$.\footnote{This is why  the dipolar asymmetry in the Hubble expansion rate or in the deceleration parameter \emph{reverses} sign in going from the heliocentric frame to the CMB frame, as shown later in Figs. \ref{fig:shell2} and \ref{fig:shell1}.} 

For comparison with previous results \cite{Brout:2022vxf}, we also make corrections for \emph{both} our motion as well as that of the SNe~Ia host galaxies (which have a peculiar redshift $z_\text{p}$), to obtain the redshifts required to construct the Hubble diagram:
\begin{equation}
\label{eqzhd}
z_\text{HD} = \frac{1+z_\text{CMB}}{1+z_\text{p}} - 1,
\end{equation}
Note that this is just the usual redshift $z$ of FLRW models in the CRF, in which the Hubble expansion of galaxies should be isotropic. However the recent finding that the dipole anisotropy in the sky distribution of  distant radio sources and quasars does \emph{not} match that expected due to our peculiar motion, as reflected in the CMB dipole \cite{Secrest:2022uvx, Dam:2022wwh,Wagenveld:2023kvi}, casts doubt on whether this procedure for `isotropising' the data is in fact valid.

\subsection{Anisotropy in the Hubble expansion rate}  

We investigate first the anisotropy in $H_0$ by parameterising it as: 
\begin{equation}
\label{H0aniso}
H = H_\text{m} + \bm{H}_\text{d}\cdot\hat{n}
\end{equation}
where $H_\text{m}$ is the monopole and $\bm{H}_\text{d}$ is the dipole component.
We fit this to data in the redshift range $z=0.023-0.15$ with 487 supernovae for analysis C1, and 480 supernovae for analysis C2. (The remaining 7 supernovae make the covariance matrix for analysis C2 non-`positive semi-definite', hence have been excluded from the analysis.) 
All cosmological and light-curve parameters ($q_0$, $j_0 -\Omega_k$, $\alpha$, $\beta$, $x_{1,0}$, $c_0$, $\sigma_{x_1}$ and $\sigma_c$) are fixed, and we estimate only $H_\text{m}$ and $\bm{H}_\text{d}$ together in direction $\hat{n}$, in both analyses C1 and C2. 

In Tables \ref{tab:HD_fix_C1} and \ref{tab:HD_fix_C2}, the cosmological parameters (and the Phillips-Tripp light curve standardisation parameters, in case of analysis C2) are fixed to two sets of values:

\begin{trivlist}
\item (i) From the Pantheon+ analysis over the whole redshift range \cite{Brout:2022vxf},
\item (ii) The standard flat $\Lambda$CDM model values ($q_0=-0.53, j_0-\Omega_k=1$).

\end{trivlist}

\begin{table}[ht!]
\centering
\begin{tabular}[H]{|c|c|c|c|c|c|c|c|c|c|} 
  \hline
  &\textbf{Frame} & $q_\text{m}, j_0$ & $H_\text{m}$ & ${H}_\text{d}$ & $l_\text{d}$ & $b_\text{d}$ & $\Delta$LLH$|_{\bm{H}_\text{d}=0}$ & $\alpha$ \\
  & & set at & (km~s$^{-1}$Mpc$^{-1}$) & (km~s$^{-1}$Mpc$^{-1}$) & (deg) & (deg) & & \\
  \hline\hline
  i) & Hel & {-0.049, -0.94} & $70.6^{+0.3}_{-0.4}$ & $-2.0^{+0.9}_{-0.9}$ & 188.8 & 36.6 & 18.1 & 3.5$\sigma$ \\
  ii) & Hel & {-0.53, 1} & $71.6^{+0.3}_{-0.3}$ & $-2.3^{+0.7}_{-0.7}$ & 187.7 & 41.4  & 34.9 & 5.3$\sigma$ \\ \hline
  i) & CMB & {-0.24, -0.33} & $71.3^{+0.3}_{-0.3}$& $2.9^{+1.1}_{-1.1}$ & 321 & 2.3& 16.6 & 3.3$\sigma$ \\
   ii) & CMB & {-0.53, 1} & $71.9^{+0.3}_{-0.2}$ & $2.4^{+1.0}_{-1.1}$ & 321.1 & -2.4& 18.7 & 3.6$\sigma$ \\ \hline
   i) & LG & {-0.03, -1.04} & $70.4^{+0.3}_{-0.3}$ & $-2.7^{+0.8}_{-0.9}$ & 241.1
   & 27.4 &38.8 & 5.6$\sigma$ \\
   ii) & LG & {-0.53, 1} & $71.5^{+0.2}_{-0.3}$ & $-2.9^{+0.6}_{-0.7}$ & 239.3 & 32.6& 71.1 & 7.9$\sigma$\\ \hline
\end{tabular}
\caption{Estimates of $H_\text{m}$ and $\bm{H}_\text{d}$ (in direction $\hat{n}: l_\text{d},\, b_\text{d}$) in the redshift range $z=0.023-0.15$ in analysis C1 (sample and redshift-\emph{dependent} light curve standardisation) while fixing all other parameters. Here $\alpha$ denotes the statistical significance with which the `no-dipole' hypothesis is rejected according to the likelihood ratio, by Wilks' theorem (with 3 d.o.f.). }
\label{tab:HD_fix_C1}
\end{table}

\begin{table}[ht!]
\centering
\begin{tabular}[H]{ 
  |c|c|c|c|c|c|c|c|c|c| } 
  \hline
  &\textbf{Frame} & $q_\text{m}, j_0$ & $H_\text{m}$ & ${H}_\text{d}$ & $l_\text{d}$ & $b_\text{d}$ & $\Delta$LLH$|_{\bm{H}_\text{d}=0}$ & $\alpha$ \\
  & & set at & (km~s$^{-1}$Mpc$^{-1}$) & (km~s$^{-1}$Mpc$^{-1}$) & (deg) & (deg) & & \\
  \hline\hline
  i) & Hel &{0.379,-1} & $70.3^{+1.2}_{-1.0}$ & $-1.8^{+1.0}_{-0.9}$ & 174.7& 21.1 & 12.0 & 2.7$\sigma$ \\
  ii) & Hel &{-0.53, 1} & $72.1^{+1.2}_{-1.1}$ & $-2.2^{+0.9}_{-0.9}$ & 172.7 & 33.3 & 20.5 & 3.8$\sigma$ \\ 
  \hline
  i) & CMB & {0.127,-0.781} & $71.0^{+1.2}_{-1.1}$& $2.7^{+0.9}_{-0.9}$& 313.2 & 14.9& 33.5 & 5.2$\sigma$ \\
  ii) & CMB & {-0.53,1} & $72.3^{+1.2}_{-1.1}$ & $2.8^{+1.0}_{-1.0}$ & 314.0 & 6.7 & 27.3 & 4.6$\sigma$ \\ 
  \hline
  i) & LG & {0.447, -1.05} & $70.1^{+1.2}_{-1.1}$ & $-2.2^{+0.7}_{-0.8}$ & 235.5& 17.3 & 30.1 &4.8$\sigma$ \\
  ii) & LG & {-0.53,1} & $72.1^{+1.1}_{-1.2}$ & $-2.6^{+0.7}_{-0.8}$ & 232.5 & 29.3 &  44.7& 6.1$\sigma$ \\ 
  \hline
\end{tabular}
\caption{Estimates of $H_\text{m}$ and $\bm{H}_\text{d}$ (in direction $\hat{n}: l_\text{d},\, b_\text{d}$) in the redshift range $z=0.023-0.15$ in analysis C2 (sample and redshift-\emph{independent} light curve standardisation), while fixing all other parameters. Here $\alpha$ denotes the statistical significance with which the `no-dipole' hypothesis is rejected according  to the likelihood ratio, by Wilks' theorem (with 3 d.o.f.). For analysis C2, we have set $\sigma_{M_0}=0.135$ which is the best-fit in the range $z=0.023-0.15$; if we instead set $\sigma_{M_0}=0.2$ (which is the best-fit in the full range $z=0-0.8$), then $\alpha$ is 1.8, 3.5 and 3.2 respectively, for the heliocentric, CMB and Local Group frame.}
\label{tab:HD_fix_C2}
\end{table}

\begin{table}[ht!]
\centering
\begin{tabular}[H]{ 
  |c|c|c|c|c|c|c|c|c|c| } 
  \hline
  \textbf{Frame} & $H_\text{m}$ & ${H}_\text{d}$ & $l_\text{d}$ & $b_\text{d}$ & $q_\text{m}$ & $j_0-\Omega_k$ & $\Delta$LLH$|_{\bm{H}_\text{d}=0}$ & $\alpha$ \\
  & (km~s$^{-1}$Mpc$^{-1}$) & (km~s$^{-1}$Mpc$^{-1}$) & (deg) & (deg) & & & & \\ 
  \hline\hline
  Hel & $69.8^{+1.1}_{-1.0}$ & $-2.2^{+0.8}_{-0.7}$ & 188.4 & 34.7 & 1.08 & -21.3 & 29.8 & 4.8$\sigma$ \\
  CMB & $70.1^{+1.1}_{-1.0}$ & $2.2^{+1.0}_{-0.9}$ & 321.8 & 2.6 & 0.77 & -13.4 & 19.5 & 3.7$\sigma$ \\
  LG & $69.9^{+1.0}_{-1.0}$ & $-2.9^{+0.7}_{-0.7}$ & 238.4 & 32.6 & 1.02 & -23.4 & 65.3 & 7.6$\sigma$ \\
  \hline
\end{tabular}
\caption{Estimates of $H_\text{m}$ and $\bm{H}_\text{d}$ (in direction $\hat{n}: l_\text{d},\, b_\text{d}$) in the redshift range $z=0.023-0.15$ in analysis C1 (sample and redshift-\emph{dependent} light curve standardisation), while simultaneously fitting all other parameters. $\alpha$ is the statistical significance with which the `no-dipole' hypothesis is rejected according to the likelihood ratio, by Wilks' theorem (with 3 d.o.f.).}
\label{tab:HD_float_C1}
\end{table}

\begin{table}[ht!]
\begin{tabular}[H]{|c|c|c|c|c|c|c|c|c|c|} 
  \hline
  \textbf{Frame} & $H_\text{m}$ & ${H}_\text{d}$ & $l_\text{d}$ & $b_\text{d}$ & $q_\text{m}$ & $j_0-\Omega_k$ & $\Delta$LLH$|_{\bm{H}_\text{d}=0}$& $\alpha$ \\
  & (km~s$^{-1}$Mpc$^{-1}$) & (km~s$^{-1}$Mpc$^{-1}$) & (deg) & (deg) & & & & \\
  \hline\hline
  Hel & $70.2^{+1.6}_{-1.5}$& $-2.1^{+0.9}_{-0.9}$ & 177.2 & 34.0 & 2.16 & -36.1 & 18.0&3.5 $\sigma$ \\
  CMB & $70.5^{+1.7}_{-1.5}$ & $2.6^{+1.0}_{-1.0}$ & 314.8 & 10.4  &1.35 & -27.2 & 24.5&4.3$\sigma$ \\
  LG & $70.2^{+1.8}_{-1.4}$ & $-2.7^{+0.8}_{-0.8}$ & 233.3 & 30.9 & 2.10 & -38.8 & 41.3 &5.8$\sigma$ \\
  \hline
\end{tabular}
\caption{Estimates of $H_\text{m}$ and $\bm{H}_\text{d}$ (in direction $\hat{n}: l_\text{d},\, b_\text{d}$) in the redshift range $z=0.023-0.15$ in analysis C2 (sample and redshift-\emph{independent} light curve standardisation), while simultaneously fitting all other parameters. $\alpha$ is the statistical significance with which the `no-dipole' hypothesis is rejected according to the likelihood ratio, by Wilks' theorem (with 3 d.o.f.).}
\label{tab:HD_float_C2}
\end{table}

\begin{figure}[ht!]
\centering
\includegraphics[width=0.8\columnwidth]{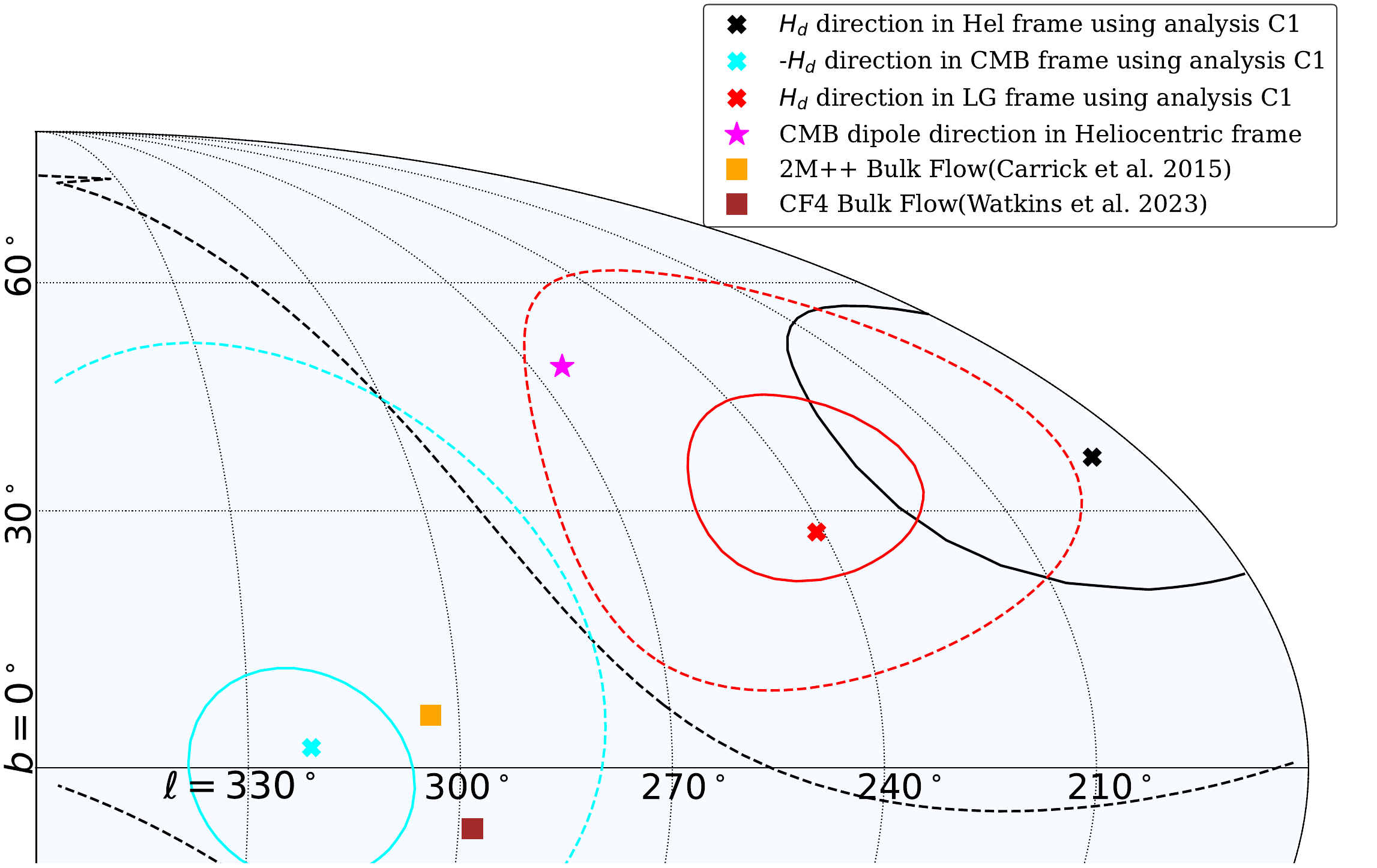}
\includegraphics[width=0.8\columnwidth]{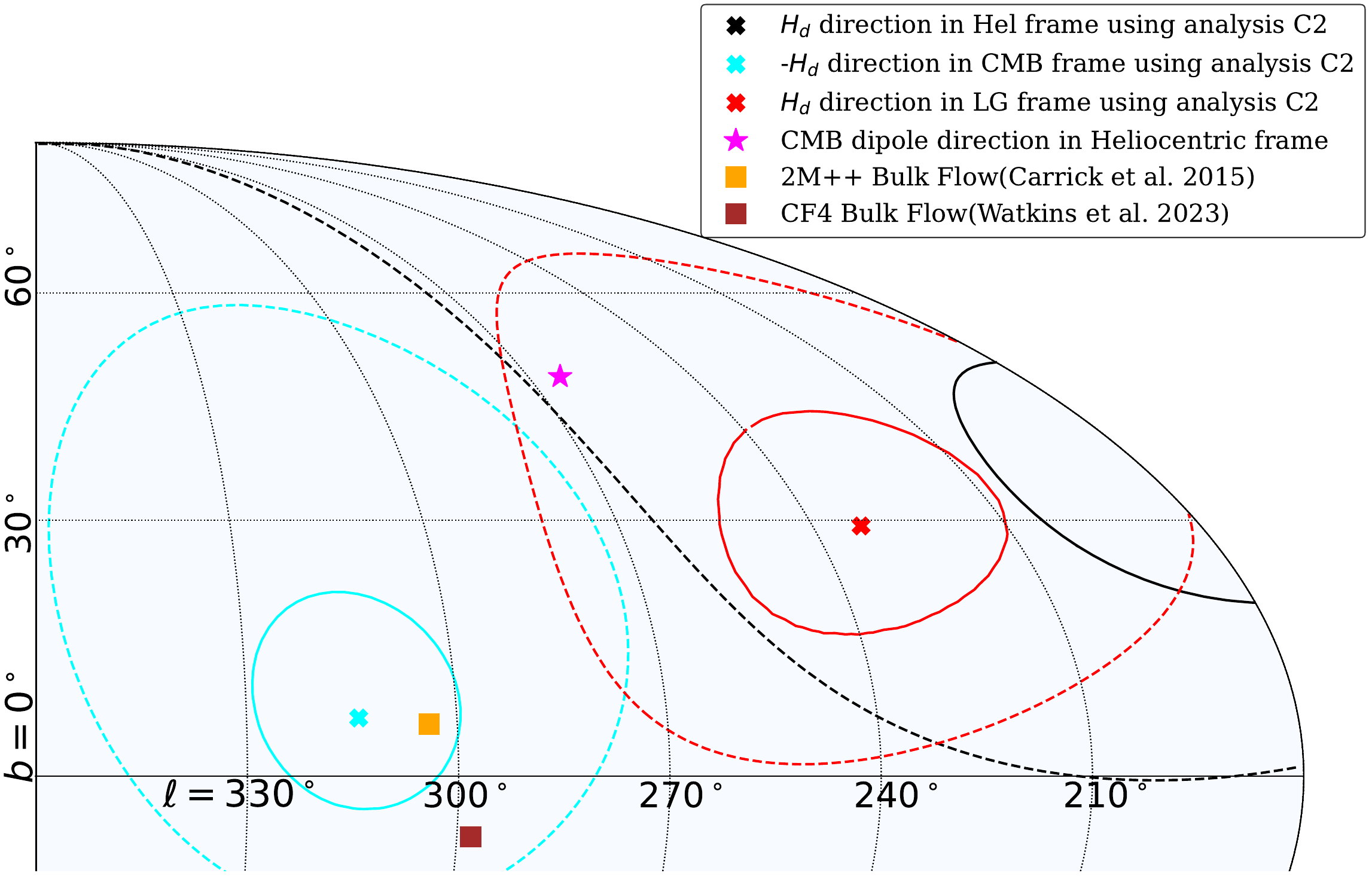}
    \caption{Mollweide view of the direction of the dipole in the Hubble parameter in Pantheon+ SNe~Ia in the redshift range $z=0.023-0.15$ for the case when cosmological parameters are fixed to $\Lambda$CDM values, in the Heliocentric, Local Group, and CMB frames. The (magenta) star denotes the CMB dipole direction. The solid and dashed lines denote $1\sigma$ and $3\sigma$ contours around the best-fit points for analyses C1 in the top panel and for analyses C2 in the bottom panel, i.e., respectively, with and without sample and redshift-dependence in the light curve standardisation. In the bottom panel, the central direction of $\bm{H}_\text{d}$ in the heliocentric frame is towards $l~\text{[deg]} = 172.7, b~\text{[deg]} = 33.3$, i.e. outside the displayed quadrant.}
\label{fig:skyH0}
\end{figure}

Tables \ref{tab:HD_float_C1} and \ref{tab:HD_float_C2} present the results when all parameters are evaluated simultaneously. Now the $\Delta \text{LLH}_{H_{\text{d}=0}}$ entries correspond to the difference between the maximum log likelihood, and its value when $\bm{H}_\text{d}$ is set to zero. Assuming that this follows a $\chi^2$ distribution (with 3 d.o.f.). Wilks' theorem \cite{Wilks:1938dza} can be employed to compute the statistical significances; we have checked that the pull distribution \cite{Nielsen:2015pga} is in fact narrower than Gaussian, so this procedure is conservative \cite{Algeri:2019lah}. Figure~\ref{fig:skyH0} shows the sky directions.

\begin{figure}[ht!]
\centering
\includegraphics[width=0.49\columnwidth]{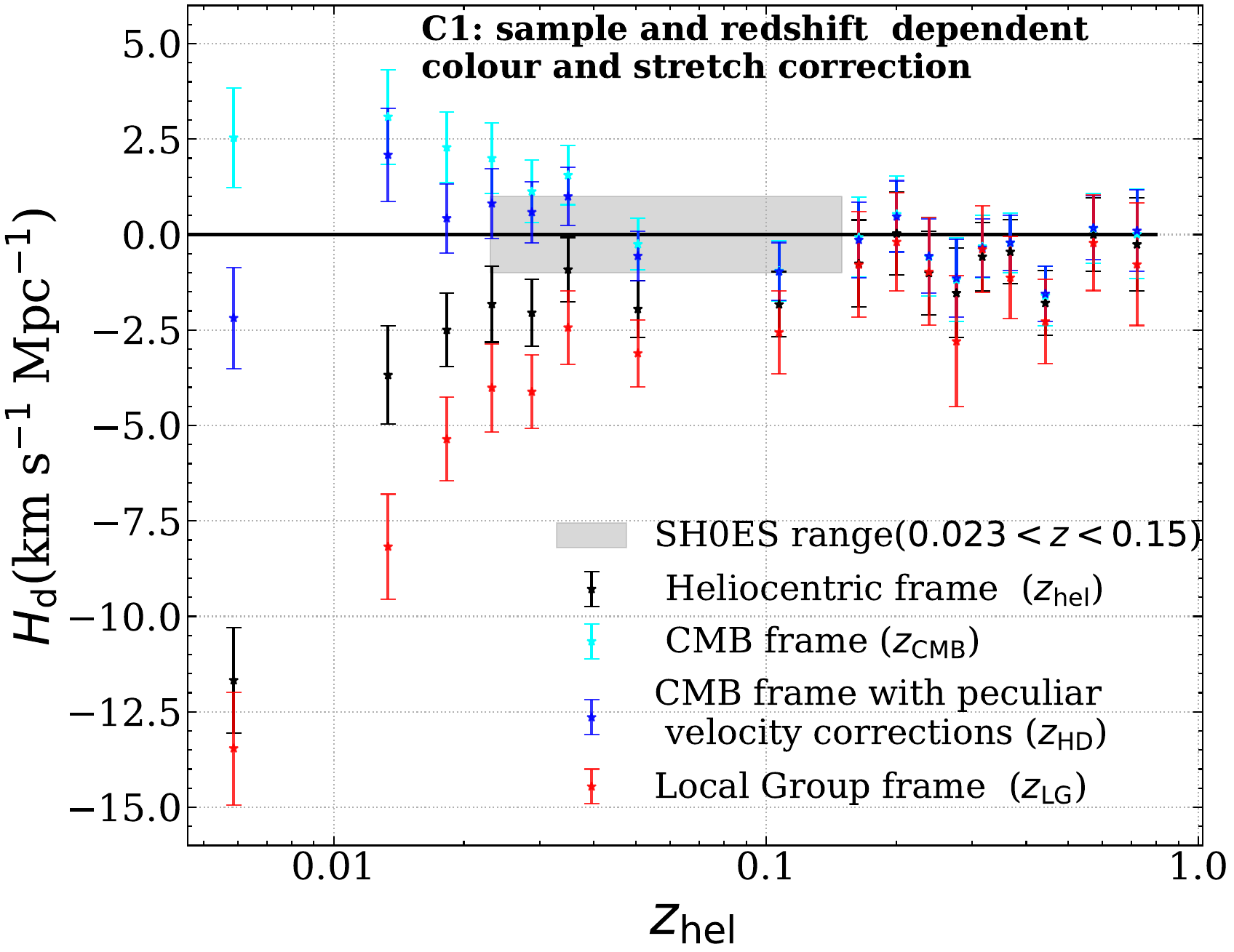}
\includegraphics[width=0.49\columnwidth]{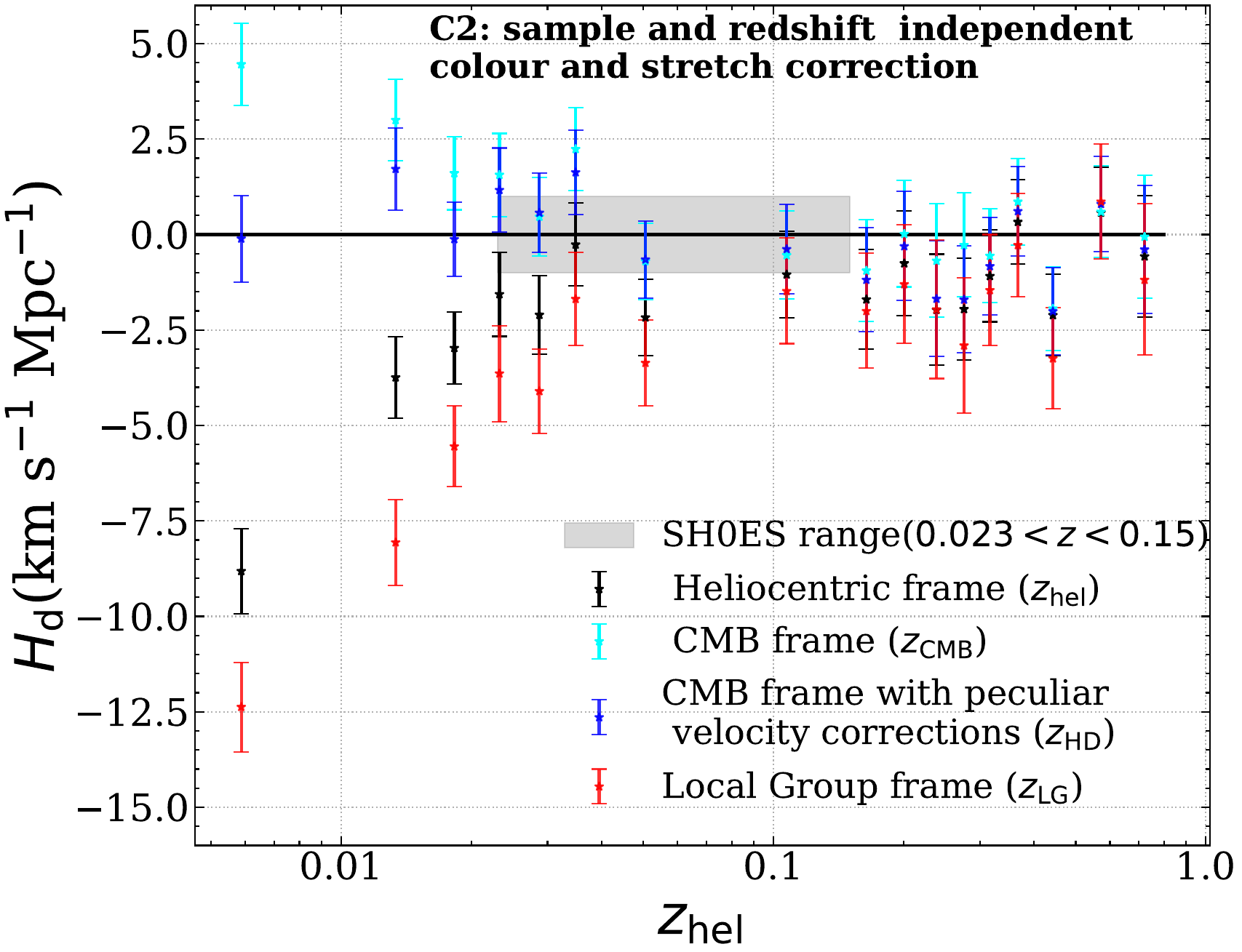}
    \caption{$\bm{H}_\text{d}$ evaluated in 17 shells (each containing around 100 supernovae) plotted against the median redshift of the shells, with all other parameters held fixed. The analyses are carried out in the heliocentric, Local Group, CMB and Hubble Diagram frames. The parametric form of the fitted dipole is scale-\emph{independent} i.e. $H = H_\text{m} + \bm{H}_\text{d}\cdot\hat{n}$, with the direction aligned with the CMB dipole. The error bars denote $\pm$1$\sigma$ uncertainties. The gray shaded region corresponds to the redshift range $z=0.023-0.15$, with its vertical width indicating the $\pm 1$~km\,s$^{-1}$\,Mpc$^{-1}$ precision on $H_0$ claimed by the SH0ES team~\cite{Riess:2021jrx}. The left and right panels correspond to analysis C1 and analysis C2 --- i.e., respectively, with and without sample and redshift-dependence in the light curve standardisation.}
\label{fig:shell2}
\end{figure}

Now we divide the full Pantheon+ sample of 1670  SNe~Ia (1653 for analysis C2) into 17 sub-samples or `shells', each containing 100 SNe~Ia (except for the last (highest redshift) shell which contains 70 for analysis C1 and 53 for analysis C2). For each shell, all cosmological and light curve correction parameters are fixed to the best-fit values obtained from the Maximum Likelihood Estimate for the full sample. Within each shell, a fit is performed to a scale-\emph{independent} dipole, with the direction fixed to the CMB dipole direction. In Fig.~\ref{fig:shell2}, we plot these dipoles in the Hubble expansion rate in each shell along with their 1$\sigma$ confidence intervals, against the median redshift of the corresponding shell.


\subsection{Anisotropy in Deceleration parameter} \label{sec:dip_in_q}

\begin{figure}[ht!]
\centering
\includegraphics[width=0.49\columnwidth]{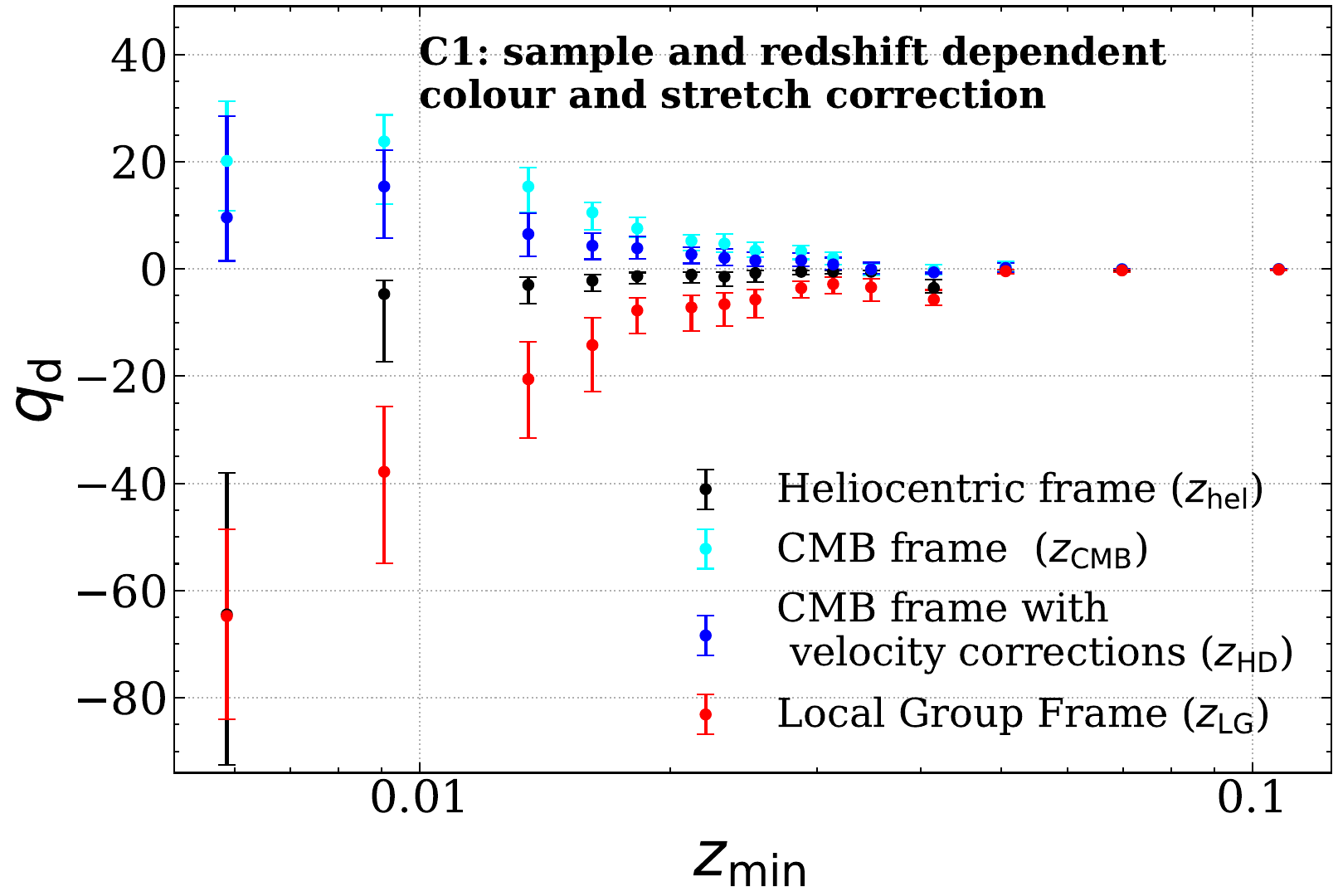}
\includegraphics[width=0.5\columnwidth]{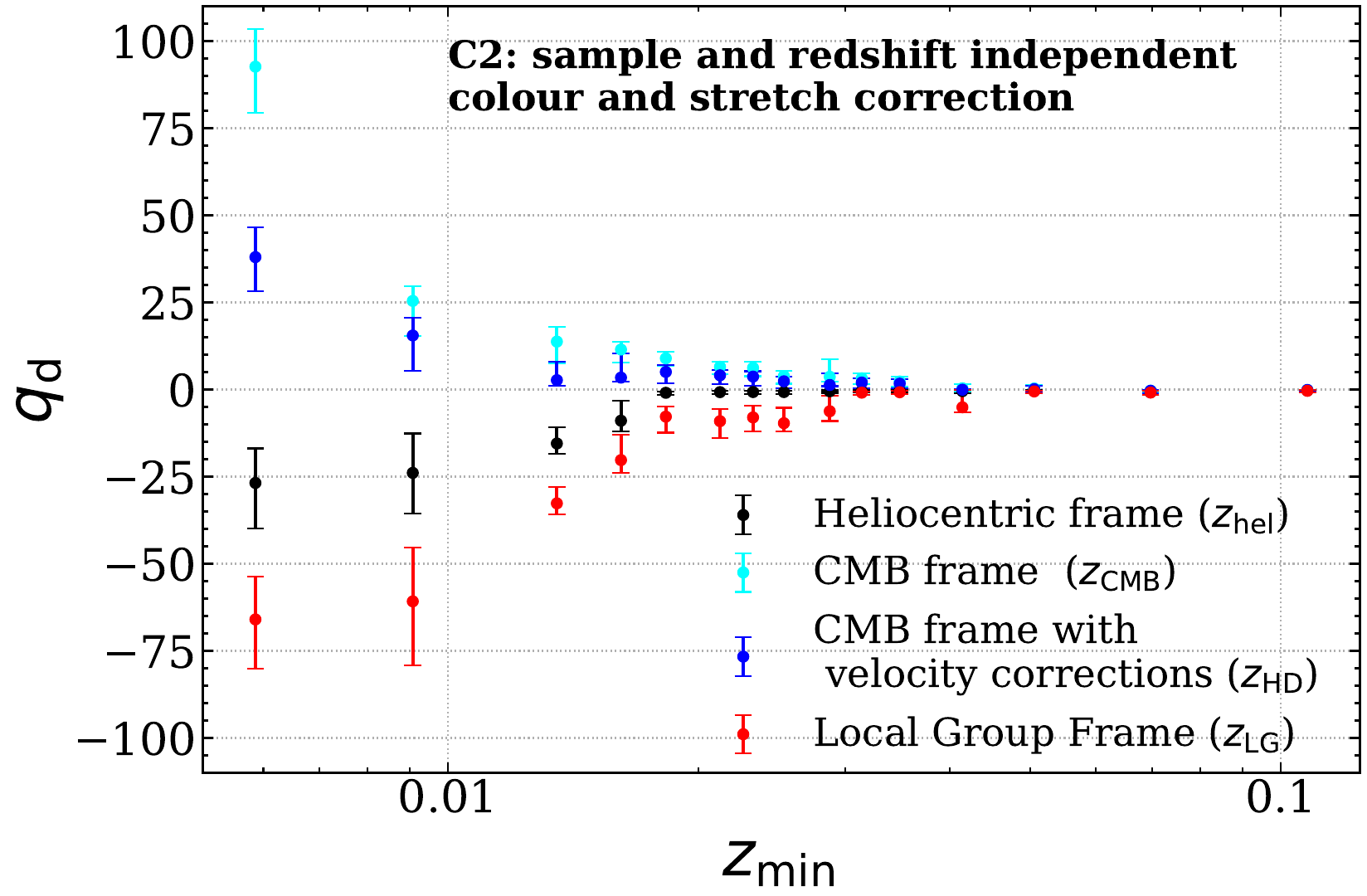}
    \caption{The scale-dependent dipole component of the deceleration parameter $\bm{q}_\text{d}$, evaluated for SNe~Ia samples with progressively higher redshift cuts ($z>z_{\text{min}}$), in the heliocentric, Local Group, CMB and Hubble Diagram frames. This is done for both sample \& redshift-\emph{dependent} (C1, left) and sample \& redshift-\emph{independent} (C2, right) lightcurve standardisation. Error bars indicate $1\sigma$ uncertainties obtained using Wilks' theorem. The heliocentric frame result is consistent  the previously reported anisotropy for JLA~\cite{Colin:2019opb}, which had a lower redshift cut at 0.00937.}
\label{fig:qdvsznn}

\end{figure}

We now look for a scale-\emph{dependent} dipolar modulation in the deceleration parameter $q_0$, adopting the same exponential decay with redshift  motivated previously~\cite{Colin:2019opb}:
\begin{equation}
q_0 = q_\text{m} + \bm{q}_\text{d} \cdot\ \hat{n} \text{e}^{-z/S} ,
\end{equation}
with $S$ being the scale on which the dipolar anisotropy dies out.
First, we fix the direction of $\bm{q}_\text{d}$ to the CMB dipole.
We progressively remove the lower redshift supernovae incrementally in steps of about 50 objects in each subsequent step, to check the dependence of the inferred cosmological parameters on the (heliocentric) redshift range of the sample.\footnote{
There are 150 duplicate entries in the Pantheon+ catalogue at $z<0.1$. When redshift cuts are applied, this can affect the count of supernovae e.g. if we apply a cut $z>0.01826$, then due to a SNe~Ia with exactly that redshift which has been recorded twice, only 249 SNe~Ia are excluded instead of 250.}  Note that the inclusion of lower redshift supernovae makes the dipole more negative, as seen in Fig.~\ref{fig:qdvsznn}.

\begin{figure}
      \centering
	   \begin{subfigure}{0.45\linewidth}
		\includegraphics[width=\linewidth]{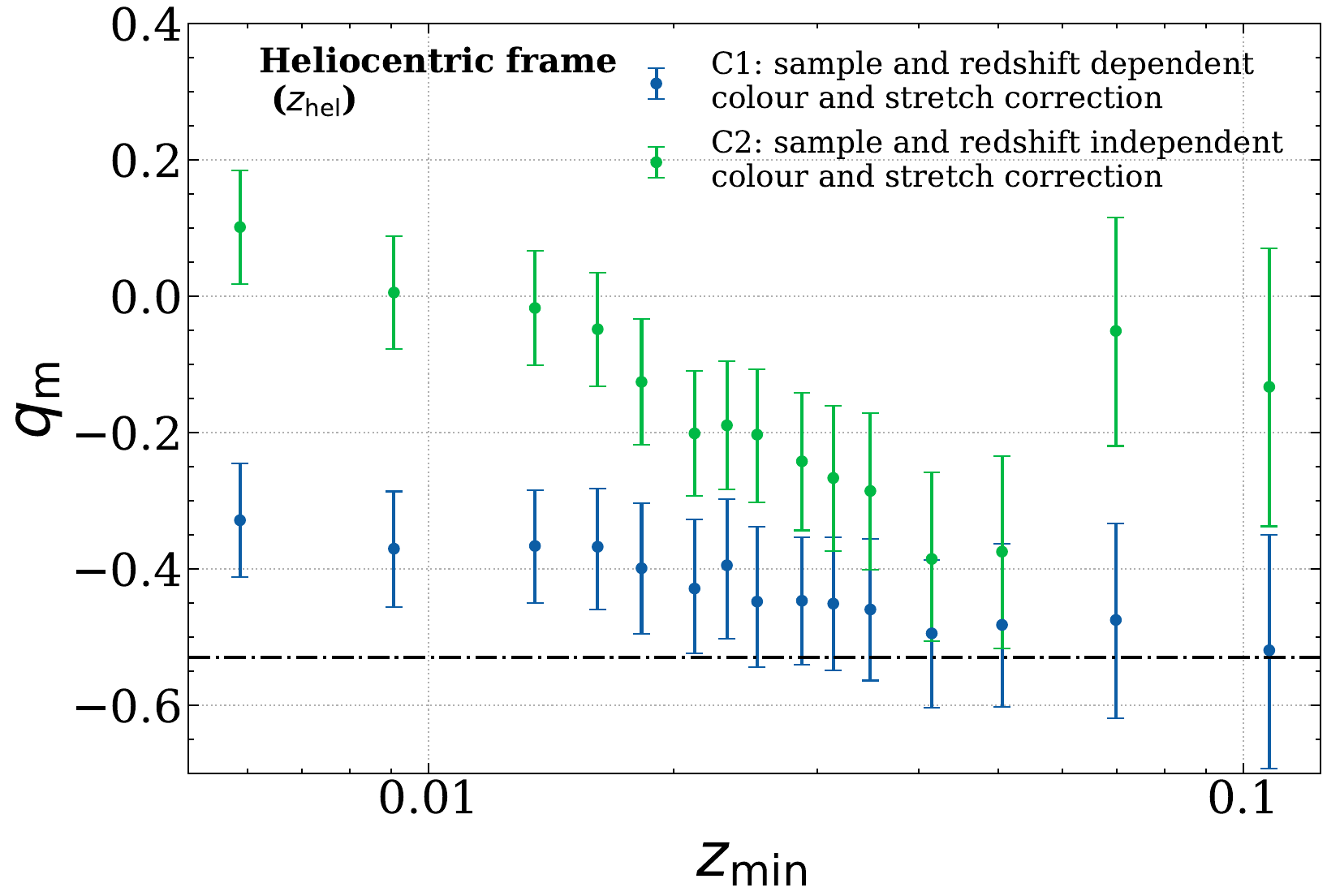}
		\label{fig:subfig1}
	   \end{subfigure}
	   \begin{subfigure}{0.45\linewidth}
		\includegraphics[width=\linewidth]{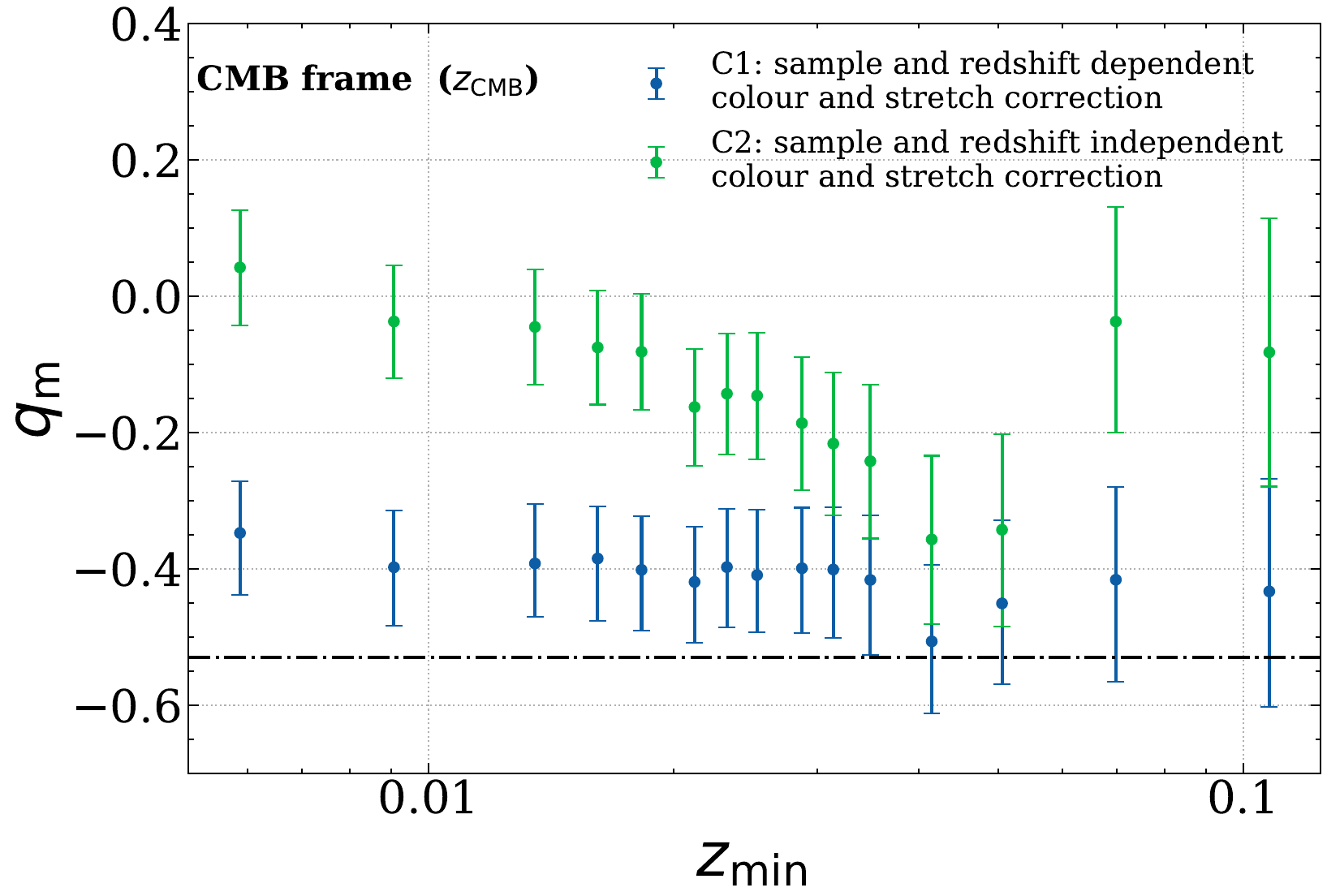}
		\label{fig:subfig2}
	    \end{subfigure}
	\vfill
	     \begin{subfigure}[b]{0.45\linewidth}
		 \includegraphics[width=\linewidth]{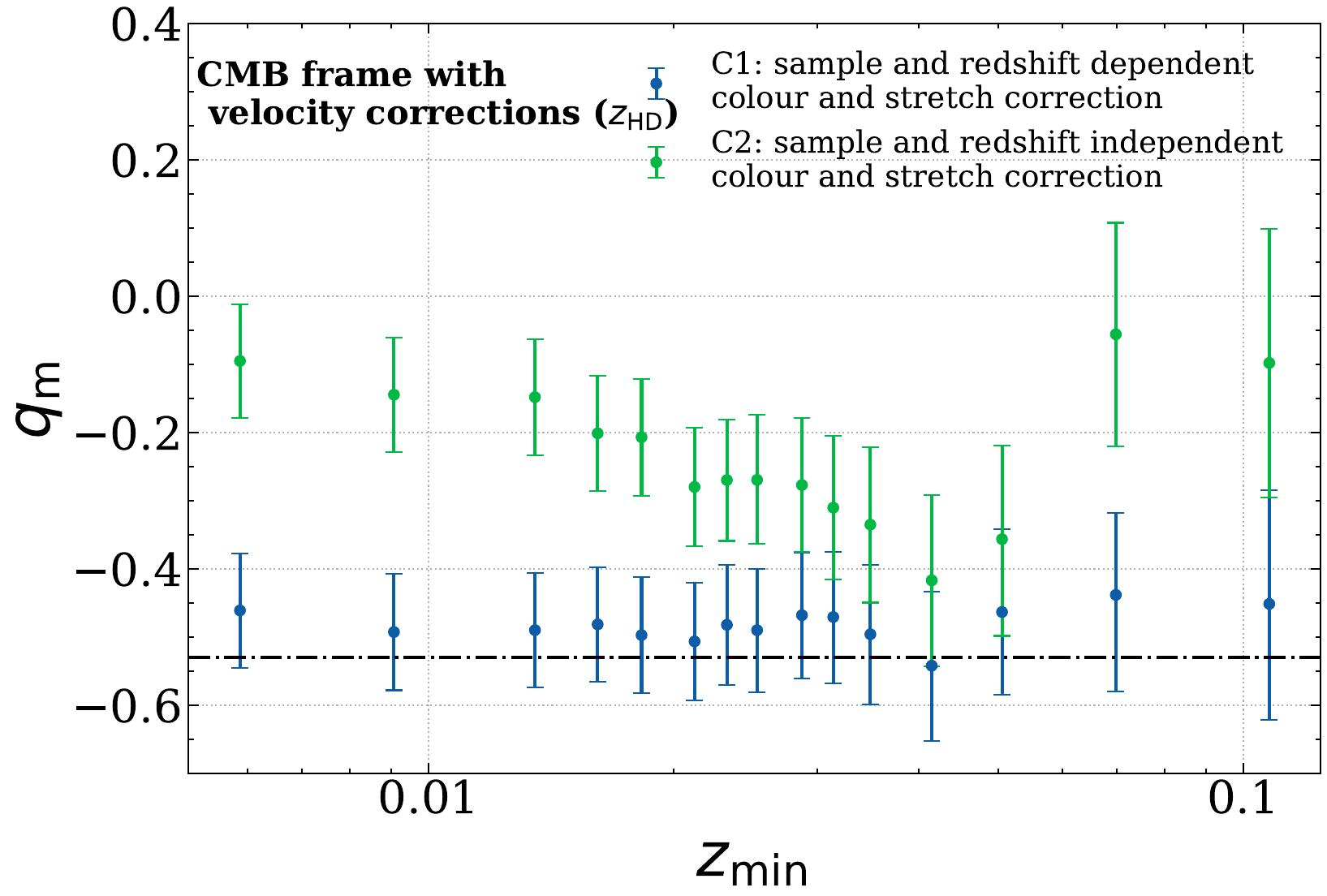}
		 \label{fig:subfig3}
	      \end{subfigure}
	       \begin{subfigure}{0.45\linewidth}
		  \includegraphics[width=\linewidth]{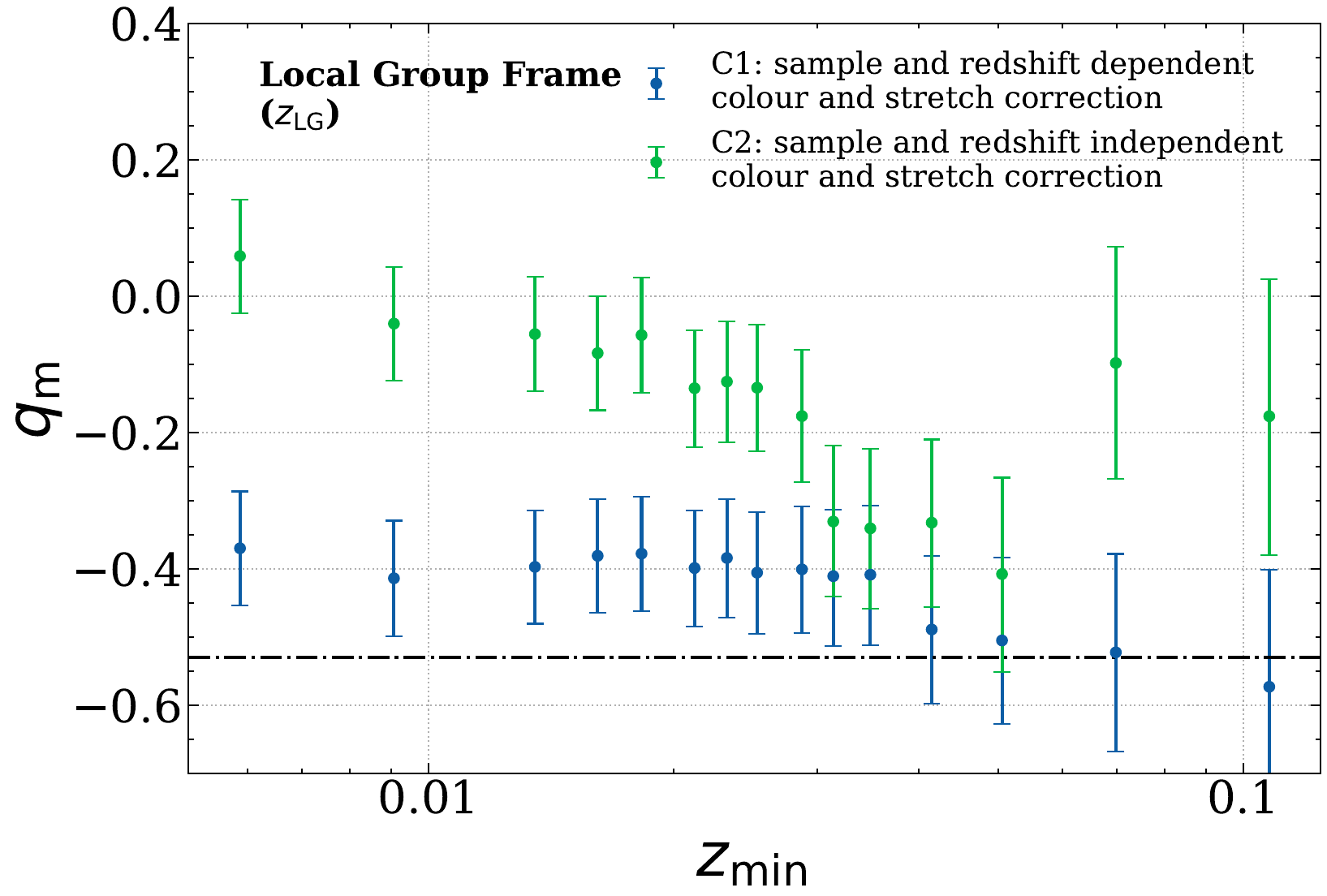}
		  \label{fig:subfig4}
	       \end{subfigure}
	\caption{The scale-independent monopole in the deceleration parameter $q_\text{m}$, for SNe~Ia samples with progressively higher cuts applied in redshift: $z>z_{\text{min}}$, in the heliocentric, Local Group, CMB and Hubble Diagram frames. Error bars indicate $1\sigma$ uncertainties obtained using Wilks' theorem. The systematically more negative values of $q_\text{m}$ obtained in analysis C1 (blue points) is because of allowing sample and redshift-\emph{dependence} in the light curve stretch ($x_1$) and colour ($c$) standardisation. In analysis C2 (green points), the averaged monopole is consistent with being zero (see Fig.~\ref{fig:scan1}). The dashed horizontal line indicates $q_0 = -0.53$, the expectation in the fiducial $\Lambda$CDM cosmology.}
	\label{fig:cumulative_qm}
\end{figure}

\begin{table}[ht!]
\setlength{\tabcolsep}{2pt} 
\begin{tabular}{|c|c|c|c|c|c|c|c|c|c|} 
  \hline
\textbf{Analysis} & \textbf{Frame} & $-2\text{log}\mathcal{L}_\text{max}$ & $\bm{q}_\text{d}$ & ${S}$  & $q_\text{m}$ & $j_0-\Omega_{k}$ & $l$ [deg], $b$ [deg]  & $\Delta$LLH$|_{\bm{q}_\text{d}=0}$ & $\alpha$ \\
   \hline \hline
 C1 & Hel & -1507 & -6.27  & 0.024 & -0.369 & 0.12  & {CMB dipole direction}  & 30.5 & 5.2 \\
 C1 & Hel & -1527 & -8.48  & 0.028 & -0.439 & 0.45 & (192.1, 38.5) & 49.9 & 6.3  \\
\hline
C2 & Hel & -184.9 & -31.8 & 0.0094& 0.01 &-0.65& {CMB dipole direction}& 36.9 & 5.7  \\
C2 & Hel & -203.6  & -34.4 & 0.011 & -0.09 &-0.39 & (188.1, 52.3) & 55.5 & 6.7 \\
\hline \hline
 C1 & LG  & -1501 & -39.6 & 0.012 & -0.411 & 0.29 & {CMB dipole direction} & 119.7 & $>7$ \\
 C1 & LG  & -1514 & -32.3 & 0.014 &-0.438  & 0.42  & (247.1, 28.1)& 132.6 & $>7$ \\
\hline
C2 & LG  & -187.2  &-61.4  & 0.0099 &-0.038  &-0.56 &{CMB dipole direction} &135.5 & $>7$   \\
C2 & LG  & -201.0  &-64.2  & 0.0099 &-0.01   &-0.39 & (248.21, 33.63) & 149.3 & $>7$ \\
\hline \hline
 C1 & CMB & -1518 & 20.5  & 0.011 &-0.39  & 0.26  & {CMB dipole direction} & 28.3 & 5.0 \\
 C1 & CMB & -1532  & 17.0  & 0.0155&-0.455  & 0.57 & (312.2, 18.9) & 42.4 & 5.7 \\
\hline
C2 & CMB & -188.3  & 21.1  & 0.011 &-0.03  &-0.51 &{CMB dipole direction} & 31.8 &5.3  \\
C2 & CMB & -168.1   & 22.5  & 0.014&-0.13  &-0.2 & (303.3,10.8)  & 51.3 & 6.4\\
\hline
\end{tabular}
\caption{Best-fit values for the anisotropy in $q_0$, in the heliocentric, Local Group and CMB frames, and with a redshift cut $z>0.00937$, both with (C1) and without (C2) sample and redshift-dependence in the light curve standardisation. The fits are better when the direction of the dipole is left free to be determined, rather than fixed to the CMB dipole direction. $\alpha$ denotes the statistical significance with which the `no-dipole' hypothesis is rejected according to the likelihood ratio, using Wilks' theorem with 2 d.o.f. when the dipole direction is fixed, and 4 d.o.f. when it is left free.}
\label{tab:qd_table}
\end{table}

\begin{figure}[ht!]
    \centering
    \includegraphics[scale=0.27]{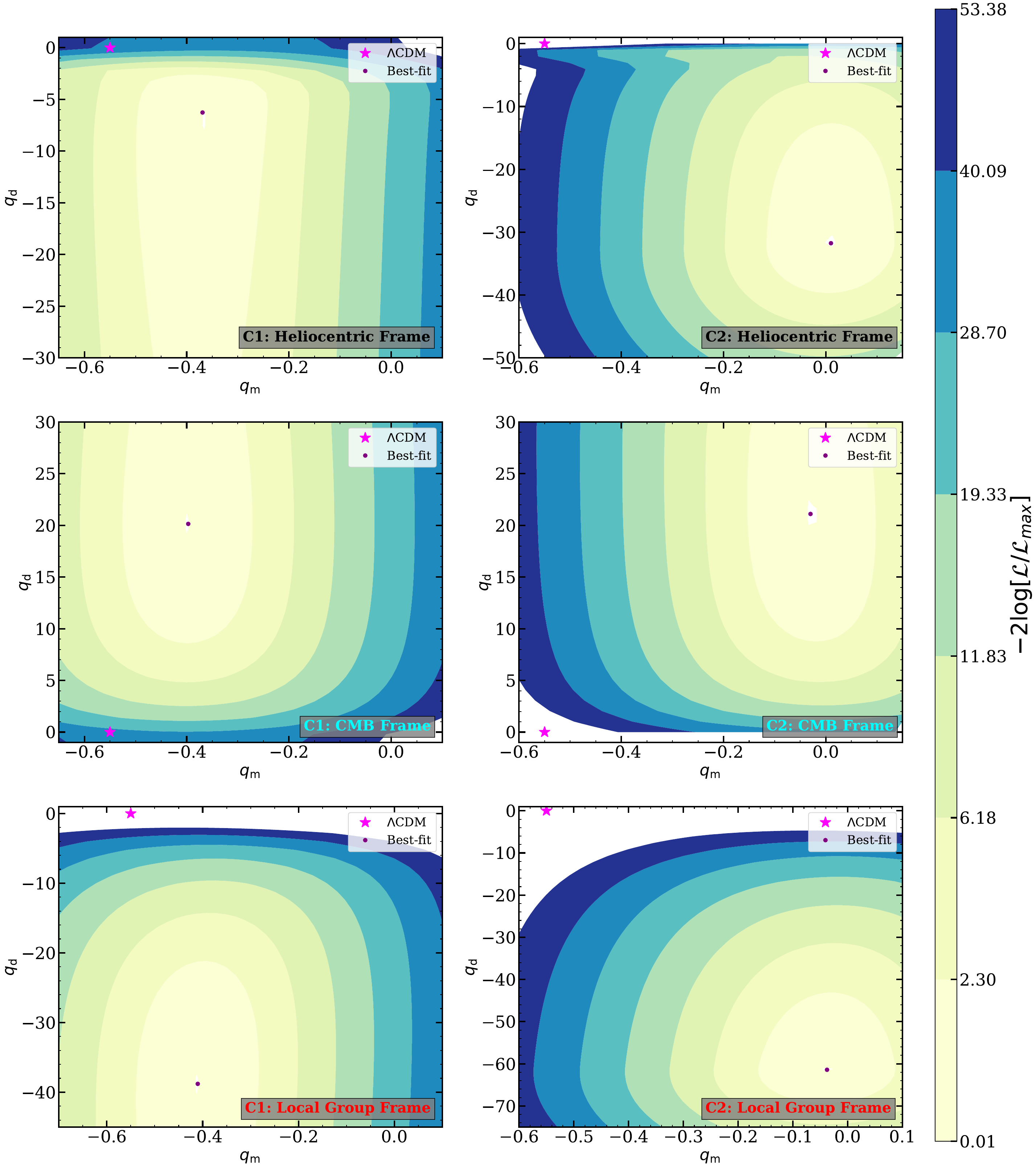}
    \caption{Contours at 1, 2, 3, 4, 5, 6 and 7 $\sigma$ for $q_\text{m}$ and $\bm{q}_\text{d}$ in the heliocentric, CMB and LG frames,  after applying a redshift cut of $z>0.00937$. The purple dot represents the best-fit values, and the magenta star denotes the expectation in the $\Lambda$CDM model. The left column is for analyses C1, while the right column is for analyses C2, i.e., with and without sample and redshift dependence in the light curve standardisation.}
    \label{fig:scan1}
\end{figure}

Specifically, with the redshift cut $z>0.00937$ (the smallest redshift in JLA), we examine the 1, 2, \ldots 5$\sigma$ contours around the best-fit parameters, $\bm{q}_\text{d}$ and $q_\text{m}$, when all other parameters are profiled over. These contours are obtained using Wilks' theorem, assuming 2 d.o.f. It is seen in Fig.~\ref{fig:scan1} that the standard $\Lambda$CDM values lie outside the $5\sigma$ contour for both analysis C1 (using magnitudes already corrected with sample- and redshift-\emph{dependent} Phillips-Tripp standardisation) and analysis C2 (incorporating Phillips-Tripp standardisation with no sample or redshift dependence as part of the fit). This confirms at higher significance the previous result concerning anisotropy of $q_0$ using JLA~\cite{Colin:2019opb}, and moreover demonstrates its frame independence.

\begin{figure}[ht!]
    \centering
    \includegraphics[width=0.8\columnwidth]{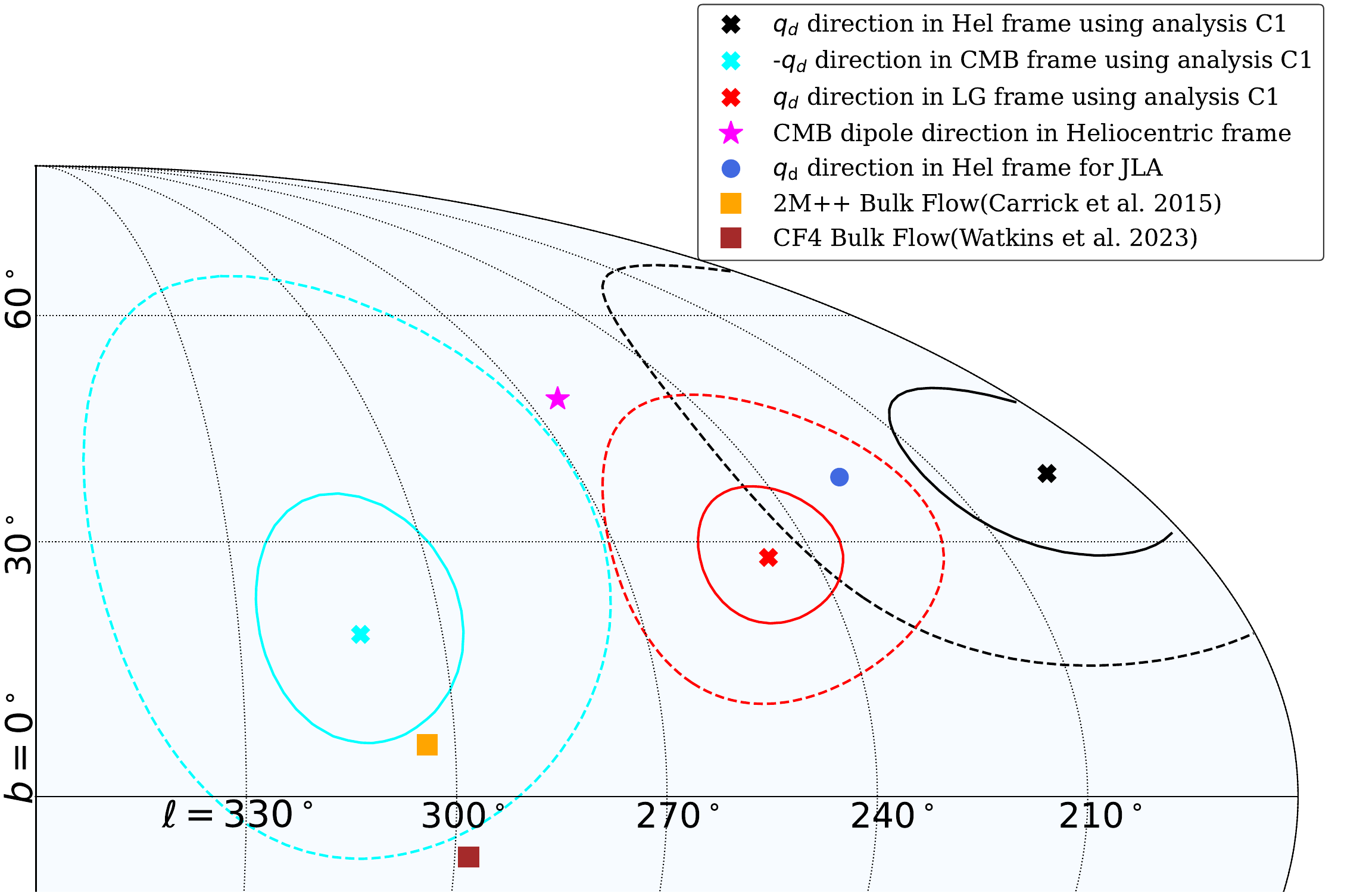}
    \includegraphics[width=0.8\columnwidth]{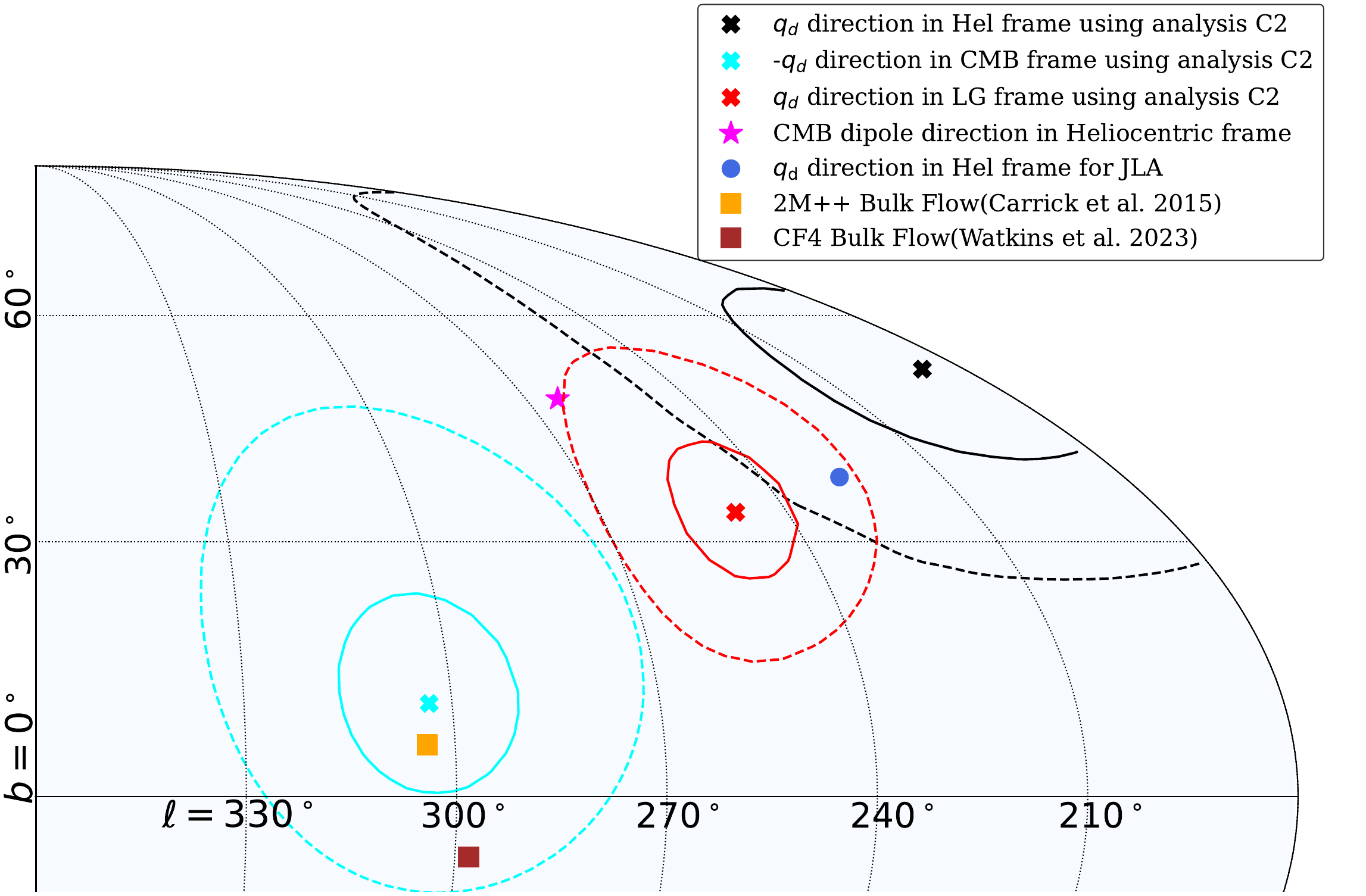}
    \caption{Mollweide view of the direction of the dipole in the deceleration parameter inferred from Pantheon+ SNe~Ia, in the heliocentric, Local Group and CMB frame.The (magenta) star denotes the direction of CMB dipole. The solid and dashed lines denote $1\sigma$ and $3\sigma$ contours around the best-fit points for analysis C1 in the top panel and for analysis C2 in the bottom panel, i.e., respectively, with and without sample and redshift-dependence in the light curve standardisation.}
    \label{fig:skyview_dipole}
\end{figure}

For the same redshift cut of $z>0.00937$, we evaluate the dipole in the LG and CMB frames as well (Table~\ref{tab:qd_table}).\footnote{We have checked that our conclusions are unaltered even when we include SNe~Ia at redshift $z > 0.8$.}
We also extend our analysis to evaluate the dipole direction (with a redshift cut $z>0.00937$), in the heliocentric frame, CMB frame, and the Local Group frame. Fig.\ref{fig:skyview_dipole} shows the best-fit directions, along with the previous best fit from Ref.~\cite{Colin:2019opb} and also the CMB dipole direction. 

\begin{figure}[ht!]
\centering
\includegraphics[width=0.495\columnwidth]{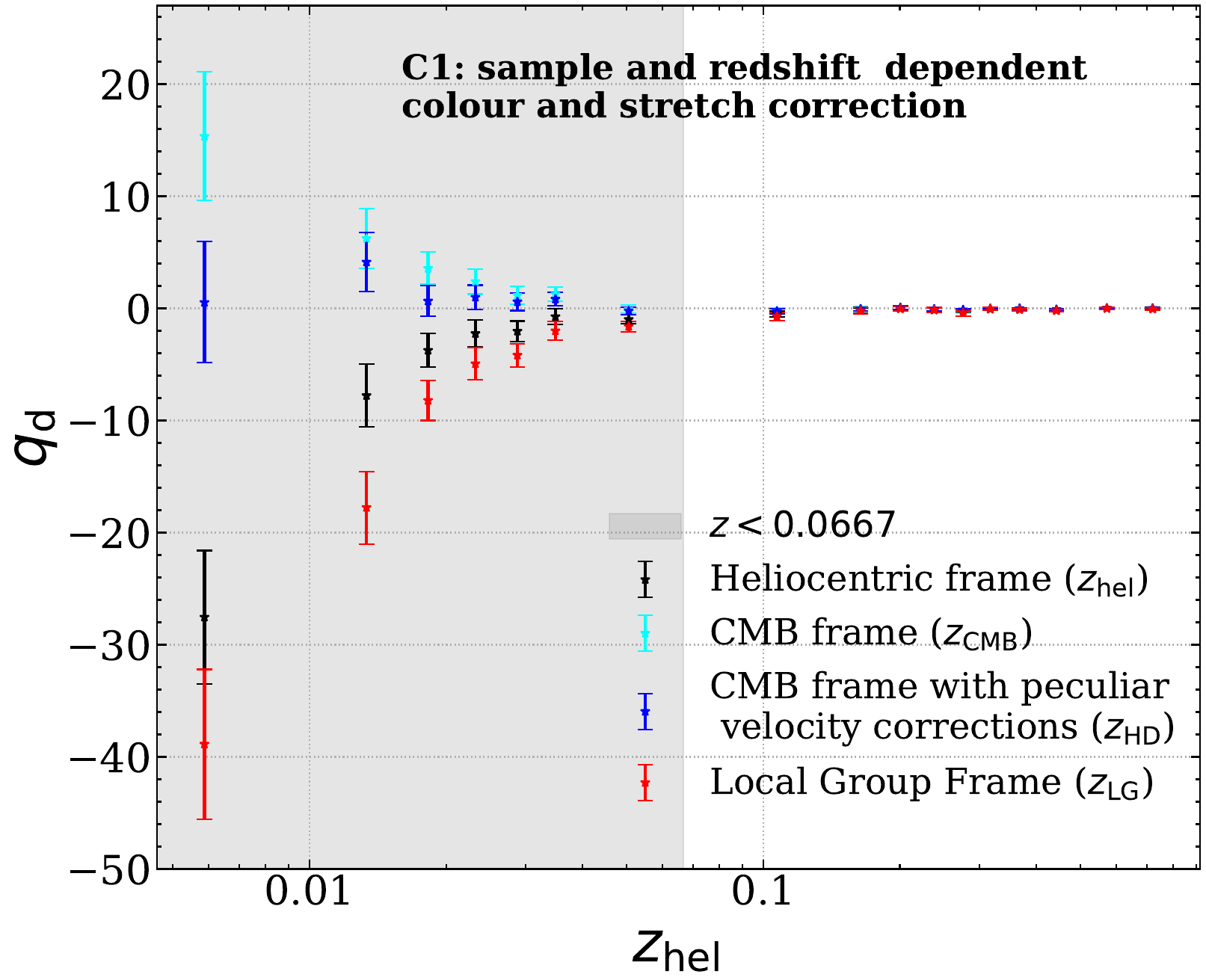}
\includegraphics[width=0.495\columnwidth]{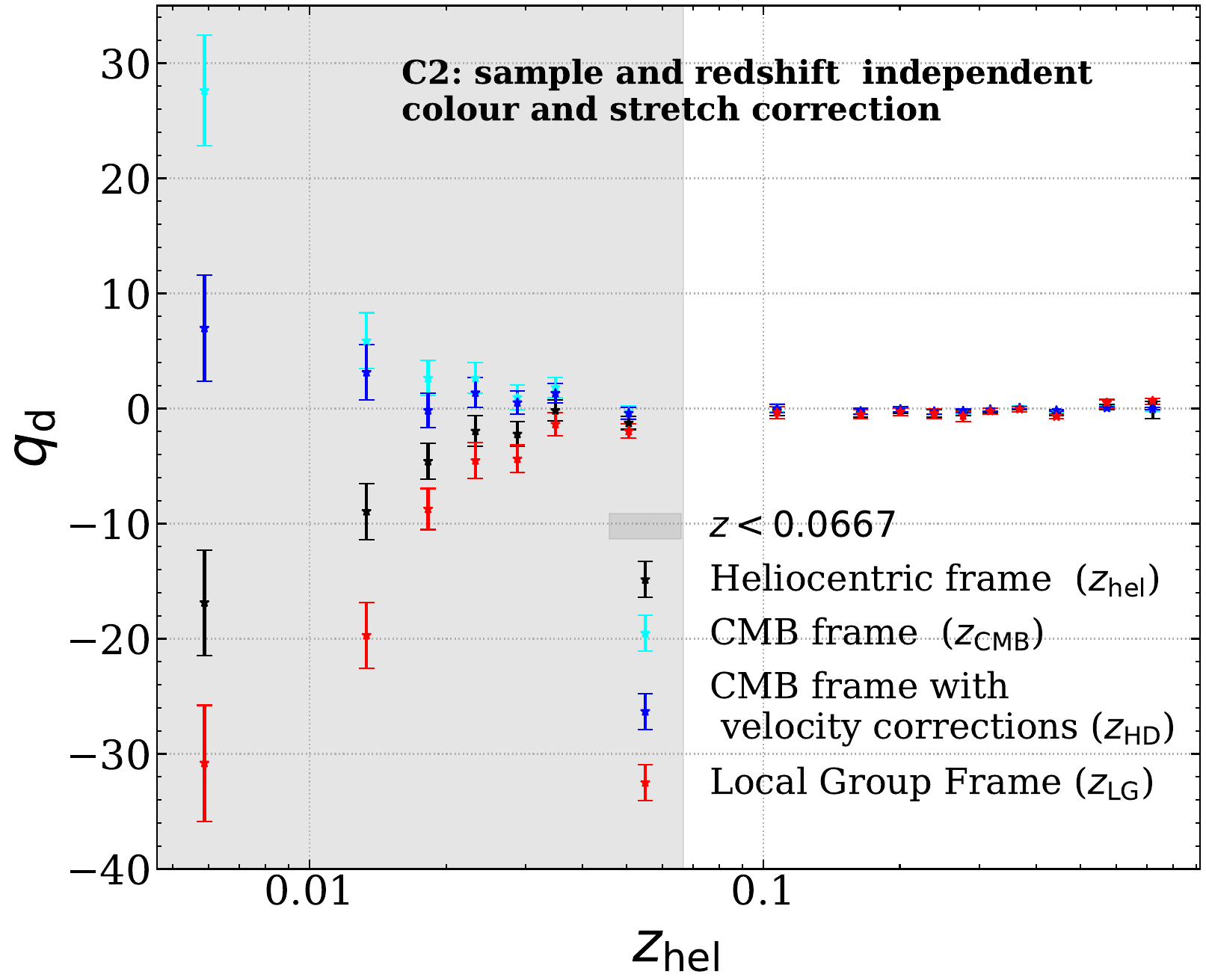}
    \caption{Scale-independent $\bm{q}_\text{d}$ evaluated in 17 shells each containing around 100 supernovae, with all other parameters held fixed, plotted against the median redshift of the shells. The analyses are done in heliocentric, Local Group and CMB frames, with the direction fixed to the CMB dipole. The gray shaded region corresponds to $z<0.0667$ i.e. distance $<200h^{-1}$Mpc. Error bars are 1$\sigma$. The parameterisation employed is scale-independent, i.e. $q=q_\text{m} + \bm{q}_\text{d}\cdot\hat{n}$ within each shell. The left panel corresponds to analysis C1 while the right panel corresponds to analysis C2, i.e., respectively, with and without sample and redshift-dependence in the light curve standardisation. The observed decay of the dipole in the deceleration parameter with redshift is a key prediction of the tilted universe scenario as explained in \S~\ref{sTCs}.} 
\label{fig:shell1}
\end{figure}

Similarly to the shell analysis for $\bm{H}_\text{d}$, (see \S 3.1), we now perform fits for a scale-\emph{independent} dipole in $q_0$, defined as:
\begin{equation}
q_0 = q_\text{m} + \bm{q}_\text{d} \cdot \hat{n},
\end{equation}
with its direction fixed to the CMB dipole direction.
This was done separately for each of the 17 shells, in the heliocentric, Local Group, and CMB frames.
In all of these, the values of $\bm{q}_\text{d}$  decay with redshift, approaching zero for $z > 0.1$, as seen in Fig.~\ref{fig:shell1}. 

\section{Theoretical considerations} \label{sTCs}

In general relativity, observers in relative motion experience different versions of \emph{reality}. This  applies in particular to observers like ourselves, who inhabit galaxies that are moving relative to the CRF. 
We explain below how observers embedded in a bulk flow can interpret observations to infer illusory local \emph{acceleration} due to their peculiar motion, even if the expansion rate is decelerating globally. Such observers should  see a (Doppler-like) dipole in the sky distribution of the deceleration parameter $q_0$, which should decay along with the bulk flow. This is the `tilted universe' scenario, originally introduced in Refs.\cite{Tsagas:2009nh,Tsagas:2011wq} and subsequently refined and extended~\cite{Tsagas:2015mua,Tsagas:2021ldz,Tsagas:2021tqa}. 

\subsection{Bulk flows and the peculiar flux}\label{ssBFPF}

In the presence of peculiar flows, the cosmic fluid is no longer perfect. The imperfection appears as an energy-flux vector ($q_a$) due to the peculiar motion \cite{ellis2012relativistic}. On an FLRW background with zero pressure, the linear `peculiar flux' is $q_a=\rho v_a$, where $\rho$ is the density of the matter and $v_a$ is the peculiar velocity perturbation.

A bulk flow is matter in motion and its flux also contributes to the relativistic gravitational effects. Then, for zero pressure, the energy and momentum conservation laws linearise to, respectively:
\begin{equation}
\dot{\rho}= -3H\rho- \text{D}^a q_a, \qquad \text{and} \qquad A_a= -\frac{1}{\rho}\,\left(\dot{q}_a+4Hq_a\right) = -\dot{v}_a- Hv_a\,.  
\label{lcls}
\end{equation}
(See Refs.~\cite{Tsagas:2007yx,ellis2012relativistic} for the nonlinear expressions.)
Recall that $q_a=\rho v_a$ to 1st order and $\dot{\rho}=-3H\rho$ to 0th order. Also, $H$ is the Hubble parameter and $\text{D}_a$ is the spatial derivative operator. According to Eq.~(\ref{lcls}b), the peculiar flux ($q_a$) implies, even in the absence of pressure, a non-zero `peculiar 4-acceleration' given by the spatial gradient of Eq.~(\ref{lcls}a):
\begin{equation}
A_a= -\frac{1}{3H}\,{\rm D}_a\vartheta+ \frac{1}{3aH}\left(\dot{\Delta}_a+\mathcal{Z}_a\right)\,,  \label{Aa}
\end{equation}
where $\vartheta={\rm D}^av_a$ and $a=a(t)$  is the scale factor~\cite{Tsagas:2015mua,Tsagas:2021ldz,Tsagas:2021tqa}. Here, $\Delta_a$ and $\mathcal{Z}_a$ describe inhomogeneities in the density and in the universal expansion respectively~\cite{Tsagas:2007yx, ellis2012relativistic}.

\subsection{Expansion and deceleration tensors}\label{ssEDTs}

The peculiar 4-acceleration can  affect the way relatively moving observers interpret data; in particular it can reverse the sign of the inferred deceleration parameter $q_0$. The range of the effect is determined by the speed and the extent of the bulk flow; since our local flow extends out to several hundred Mpc, this can create the illusion of recent \emph{global} acceleration~\cite{Tsagas:2015mua,Tsagas:2021ldz}. Observers should then see the characteristic signature of peculiar motion in the data, namely a (Doppler-like) dipole in $q_0$~\cite{Tsagas:2009nh,Tsagas:2011wq}. To the `tilted observer' in the bulk flow, the expansion rate should appear to accelerate faster along a particular direction in the sky and equally slower in the antipodal direction. This is indeed what  the SNe~Ia data shows (see Fig.~\ref{fig:shell1}).

To see why this is expected to be a local effect which should fade away on large scales, consider the familiar expansion tensor~\cite{Tsagas:2007yx,ellis2012relativistic}
\begin{equation}
\Theta_{ab}= \frac{1}{3}\,\Theta h_{ab}+ \sigma_{ab}\,,  \label{Thetaab}
\end{equation}
using which we introduce the deceleration tensor:
\begin{equation}
Q_{ab}= -\left(h_{ab}+\frac{9}{\Theta^2}\, h_a{}^ch_b{}^d\dot{\Theta}_{cd}\right)\,.  \label{Qab}
\end{equation}
Here, $h_{ab}$ is the 3-metric, $\Theta$ is the expansion scalar (with $\Theta=3H$) and $\sigma_{ab}$ is the shear tensor. The first tracks anisotropy in the expansion rate, while the second tracks anisotropy in its rate of change. The traces of Eq.~(\ref{Thetaab}) and Eq.~(\ref{Qab}) are, respectively, $\Theta_a{}^a=3H$ and $Q_a{}^a=3q$, where $q = -[1+(3\dot{\Theta}/\Theta^2)]= -[1+(\dot{H}/H^2)]$ is the familiar deceleration parameter (we drop the usual subscript $_0$). Suppose now that $u_a$ and $\tilde{u}_a$ are the 4-velocities of two observers in relative motion, with $\tilde{u}_a=u_a+v_a$. Let us identify the former observer with the CRF and the latter with a typical bulk flow with peculiar velocity $v_a$.\footnote{Tilde denote variables in the matter (i.e. bulk flow) frame, with their non-tilded counterparts evaluated in the CRF. We consider the LG frame to be closest to this frame.} To linear order, the difference between $Q_{ab}$ and $\tilde{Q}_{ab}$ is entirely due to the observers' relative motion, namely:
\begin{equation}
\tilde{Q}_{ab}- Q_{ab}= 2qu_{(a}v_{b)}- \frac{1}{H^2}\left({\rm D}_{(b}v_{a)}\right)^{\cdot}\,.  \label{tQvsQ}
\end{equation}

\subsection{Bulk flows and apparent acceleration}\label{ssBFAA}

By taking the trace of Eq.~(\ref{tQvsQ}) using Eq.~(\ref{lcls}b) and Eq.~(\ref{Aa}), while keeping in mind that $|\vartheta|/H\ll1$ at the linear level and taking into account the near flatness of the universe, we arrive at (restricting to sub-Hubble scales):
\begin{equation}
\tilde{q}- q= \frac{1}{3H^2}\,{\rm D}^aA_a= -\frac{1}{9H^3}\,{\rm D}^2\vartheta\,,  \label{tq1}
\end{equation}
with $\tilde{q}$ and $q$ representing the deceleration parameters measured by the bulk-flow observers and in the CRF respectively~\cite{Tsagas:2015mua,Tsagas:2021ldz}. Therefore, $\tilde{q}\neq q$, solely due to the peculiar motion of the matter. The spatial Laplacian on the r.h.s. above implies scale-dependence. After a simple harmonic decomposition we obtain:
\begin{equation}
\tilde{q}= q+ \frac{1}{9}\left({\frac{\lambda_H}{\lambda}}\right)^2 \frac{\vartheta}{H}\,,  \label{tq2}
\end{equation}
where $\lambda_H$ is the Hubble radius and $\lambda$ the scale of the bulk-flow. Expressions (\ref{tq1}) and (\ref{tq2}) are identical to those obtained in Refs.~\cite{Tsagas:2015mua,Tsagas:2021ldz,Tsagas:2021tqa}, but here we have provided an alternative derivation.\footnote{Although Eq.~(\ref{tq2}) has been derived here for simplicity on an Einstein-de Sitter background, it holds in all perturbed FLRW and Bianchi models~\cite{Tsagas:2021tqa}.} Accordingly, $\tilde{q}$ may have a \emph{negative} value even when $q$ is globally positive, on sub-horizon scales (where $\lambda_H/\lambda\gg1$) in a locally contracting bulk flow (where $\vartheta<0$). A recent study using the 2M++ survey data suggests that our local bulk flow is indeed \emph{contracting}~\cite{Pasten:2023tpg}.

\subsection{Bulk flows and apparent dipoles}\label{ssBFADs}

Thus if the acceleration inferred from SNe~Ia observations is an artefact of our local bulk flow, there should be a (Doppler-like) dipole in the sky-distribution of $q_0$, 
and its magnitude should decay with increasing redshift~\cite{Tsagas:2011wq}. 
Now we demonstrate how such a apparent dipole in $q_0$ emerges, entirely because of the (tilted) observer's relative motion, rather than being an intrinsic anisotropy. Evaluating the deceleration tensor in the bulk flow frame ($\tilde{Q}_{ab}n^an^b$) and in the CRF ($Q_{ab}n^an^b$), and projecting them along a fixed spatial direction ($n_a$) yields:
\begin{equation}
\tilde{Q}_{ab}n^an^b= Q_{ab}n^an^b+ \frac{1}{H}\,n^a{\rm D}_a \left[\left(v_b-\frac{1}{H}\dot{v}_b\right)n^b\right]\,,
\label{tQn1}
\end{equation}
given the linear commutation law $({\rm D}_bv_a)^{\cdot}={\rm D}_b\dot{v}_a-H{\rm D}_bv_a$. Since there is no anisotropy in the CRF, $Q_{ab}n^an^b=q$. Then, defining $V_a=v_a-\dot{v}_a/H$ and $V=V_an^a$, we get:
\begin{equation}
\tilde{Q}_{ab}n^an^b= q+n^a{\rm D}_aV, \quad \text{and} \quad \tilde{Q}_{ab}n^an^b= q-n^a{\rm D}_aV\,, 
\label{tQn2}
\end{equation}
when $V_a\uparrow\uparrow n_a$ and $V_a\uparrow\downarrow n_a$ respectively. Consequently, bulk-flow observers moving with peculiar velocity $v_a$ will assign the value (\ref{tQn2}a) to the deceleration parameter along $n_a=v_a-\dot{v}_a/H$. At the same time, observers moving the opposite way (with $\bar{v}_a=-v_a$) will measure the value (\ref{tQn2}b) along the same direction. 
Thus Eqs.~(\ref{tQn2}a) and (\ref{tQn2}b) denote a Doppler-like dipole due to the observers' peculiar motion.
Note that the direction of the $q$-dipole ($n_a$) and that of the local bulk velocity ($v_a$) may not coincide exactly; the key prediction, however, is that the dipole should decay with redshift as its magnitude depends on the spatial gradient of the velocity field as seen in Eqs.(\ref{tQn1}) and (\ref{tQn2}).

Expressions analogous to (\ref{tQn1}) and (\ref{tQn2}) apply to the expansion tensor as well. In particular, evaluating Eq.~(\ref{Thetaab}) in both the CRF and the bulk flow frame, and then comparing the linearised results, yields:
\begin{equation}
\tilde{\Theta}_{ab}n^an^b= H+ n^a{\rm D}_a \left(v_bn^b\right)= H\pm n^a{\rm D}_av,  \label{tThetan1}
\end{equation}
where the $\pm$ signs correspond to $v_a\uparrow\uparrow n_a$ and $v_a\uparrow\downarrow n_a$ respectively. Thus the bulk flow induces a (Doppler-like) dipole in the Hubble parameter too. 
As with the $q$-dipole, the magnitude of the $H$-dipole should also decay with redshift (as is in fact seen in Fig.~\ref{fig:shell2}). However since the $q$-dipole is also affected by the time-derivative of the peculiar velocity ($\dot{v}_a$), its magnitude and direction may differ from those of the $H$-dipole.

\section{Discussion} \label{discuss}

\begin{figure}[htb]
    \centering
    \includegraphics[width=1.0\columnwidth]{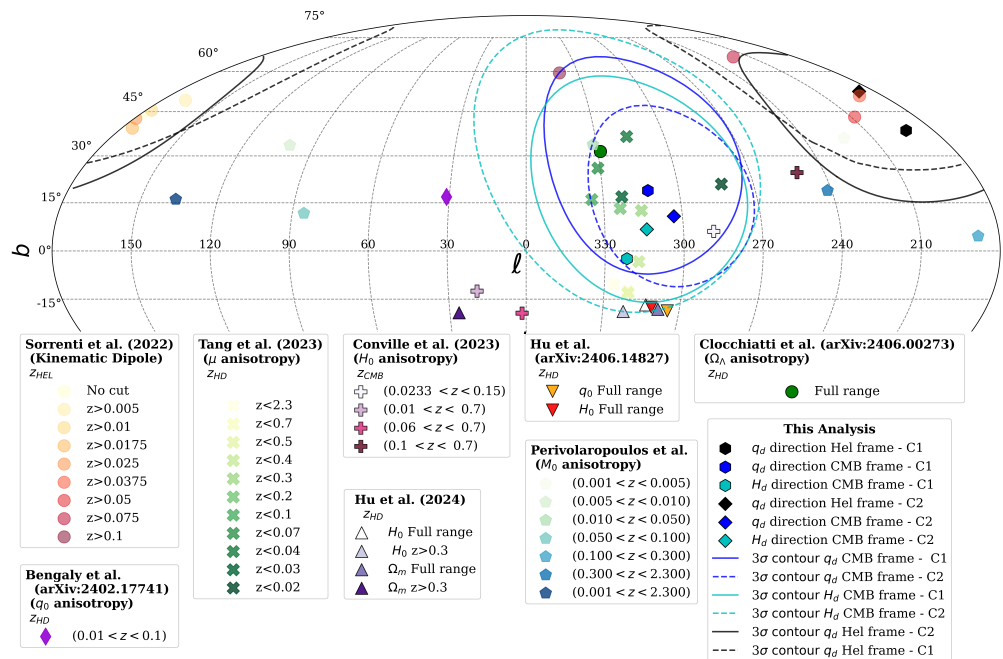}
    \caption{Mollweide view (Galactic coordinates) of all dipoles found in Pantheon+, along with the dipole directions and 3$\sigma$ contours obtained in the present analysis.}
    \label{fig:All_dipoles}
\end{figure}

To place the present work in context, we now comment on several other studies of anisotropy in the Pantheon+ catalogue~\cite{Clocchiatti:2024wxj, Hu:2023eyf,  McConville:2023xav, Hu:2024qnx, Cowell:2022ehf, Sorrenti:2022zat,Perivolaropoulos:2023tdt,Bengaly:2024ree, Tang:2023kzs}, all of whose findings are shown together in Fig.~\ref{fig:All_dipoles} for easy comparison. It should be emphasised however that the methodology, and even the data set used, varies between the authors. For a fair comparison, it is essential that only \emph{public} data available on a version controlled repository be used, as there have been instances of sudden changes in supernova data without any explanation provided as noted in Ref.~\cite{Rameez:2019wdt} regarding the Pantheon catalogue \cite{Pan-STARRS1:2017jku}.

Ref.~\cite{Clocchiatti:2024wxj} examines anisotropies in the fitted value of $\Omega_{\Lambda}$ and reports two major dipoles, one of which lies close to the North Galactic Pole-South Galactic Pole axis, while the other lies $50^{\circ}$ away from the CMB dipole and  within 3$\sigma$ of the dipole direction we find in $q_0$ (in the CMB frame). The authors believe that the first dipole may be due to anisotropic sky coverage of Pantheon+ SNe~Ia, while the second is consistent with expectations for the tilted universe~\cite{Tsagas:2015mua,Tsagas:2021ldz,Tsagas:2021tqa}, as  outlined in \S~\ref{sTCs}.

Refs.~\cite{Hu:2023eyf} and \cite{McConville:2023xav} determine the dipole in $H_0$ in the CMB frame using redshifts which also incorporate peculiar velocity corrections for the SNe~Ia host galaxies (i.e. $z_\text{HD}$); they find $H_0$ to be larger in the hemisphere encompassing the CMB dipole direction, similar to our result. Ref.~\cite{Hu:2023eyf} estimates the anisotropy in matter density $\Omega_\text{m}$ and $H_0$ using a `region fitting method', finding an anisotropy in both at $2.8\sigma$ and $4\sigma$, respectively. For the $H_0$ anisotropy, the direction of the dipole they find for the full redshift range is within (2--3)$\sigma$ of our estimated dipole direction in the CMB frame, for analysis C1 and C2 respectively; they obtain $\Delta H_0 \sim 2$~km\,s$^{-1}$\,Mpc$^{-1}$. Ref.~\cite{McConville:2023xav} estimates the anisotropy in a similar redshift range $0.0233<z<0.15$ to our analysis, and finds variations of $4~\mathrm{kms^{-1}Mpc^{-1}}$ in the Hubble parameter. Their dipole direction is within 3$\sigma$ of our estimate for the Hubble dipole (in the CMB frame) in the same redshift range. 

Ref.~\cite{Hu:2024qnx} estimates the anisotropy in both $q_0$ and $H_0$ employing Pad\'e approximants, obtaining results which are within $3\sigma$ of our estimated Hubble dipole direction. However, their estimate for $\bm{q}_\text{d}$ in the CMB frame is compatible with zero and disagrees with our result; this is likely because of their use of $z_\text{HD}$  (see comment below).

Ref.~\cite{Cowell:2022ehf} estimates the dipole in the deceleration parameter along with the quadrupole in the Hubble parameter. We see a dipole of comparable size in a similar redshift range, $z =0.023-0.8$, but find that the evidence for a quadrupole component to the Hubble expansion is significantly weaker than that for the dipole in the deceleration parameter (see Appendix \ref{sec:APPA}). Ref.~\cite{Kalbouneh:2022tfw} too had noted the absence of significant quadrupolar or higher order multipole components in the Pantheon catalogue.

The direction of the dipole in $q_0$ obtained in Ref.~\cite{Sorrenti:2022zat} is very close to ours with a redshift cut of $z=0.00937$. With $z_{\text{cut}}=0.01$, they infer the direction (141.1$^\circ$, 34.4$^\circ$) which is just  $7.1^{\circ}$ away from the dipole we find with our C1 analysis in the heliocentric frame, and $7.5^\circ$ away from our C2 analysis result, also in the heliocentric frame. 

Ref.~\cite{Perivolaropoulos:2023tdt} investigates anisotropy in the absolute magnitude of Pantheon+ supernovae using the `hemisphere comparison' Method. They employ $z_\text{HD}$ in their analysis and find consistency with isotropic Monte Carlo simulations in most redshift bins, but a sudden anisotropic transition in the lowest redshift bin at 30 Mpc. 

Ref.~\cite{Bengaly:2024ree} also tests the anisotropy of the deceleration parameter in the redshift range, $z=0.01-0.1$ using the cosmographic luminosity distance expansion to 2nd-order. They find a maximum variation of $\Delta_{q_0} = 3.06$ and conclude that $q_0$ is isotropic, again because of  using $z_\text{HD}$ in their analysis. Moreover their anisotropy direction does not align with the $\bm{q}_\text{d}$ direction from our analysis.

Ref.~\cite{Tang:2023kzs} investigates the dipole anisotropy in the distance modulus in different redshift ranges, finding a very small dipole and consistency with isotropy. However, the directions they find all lie within $3\sigma$ of the dipole we see in the Hubble parameter in the CMB frame.

Ref.~\cite{Sorrenti:2024ugq} analyzes Pantheon+ using cosmography and estimates a local monopole along with a dipole in the redshift. From the monopole, they infer a local infall within a sphere of radius $120h^{-1}$~Mpc with a velocity $\sim 100$~km\,s$^{-1}$, while the dipole is interpreted as due to a bulk flow of $317$~km\,s$^{-1}$ in the direction (RA, Dec)= ($204^{\circ}$, $-53^{\circ}$) close to the Shapley supercluster. Note that the local infall reported \cite{Sorrenti:2024ugq} suggests that our Galaxy resides in a \emph{contracting} bulk flow, in agreement with the study~\cite{Pasten:2023tpg} and in line with the expectation in the tilted-universe scenario~\cite{Tsagas:2015mua,Tsagas:2021ldz,Tsagas:2021tqa}.

It is worth noting that some degree of anisotropy was also found in the Pantheon dataset \cite{Pan-STARRS1:2017jku}, the predecessor to Pantheon+. Ref. \cite{Krishnan:2021jmh} found a directional dependence in $H_0$ aligned with the CMB dipole, with $\Delta H_0 \sim 1~\text{km}\, \text{s}^{-1}\text{Mpc}^{-1}$. Ref. \cite{Zhai:2022zif} demonstrated that sample variance can lead to fluctuations in $H_0$ of order $\pm1~\text{km}\,\text{s}^{-1}\text{Mpc}^{-1}$ but also found a variation of $4~\text{km}\,\text{s}^{-1}\text{Mpc}^{-1}$ between two hemispheres, with systematically higher $H_0$ in the hemisphere aligned with the CMB dipole direction. Earlier, a general method taking into account possible directional variations of SNe Ia light curve corrections had been proposed and applied to the Union~2.1 catalogue, also finding the strongest deviations from isotropy in the CMB dipole direction \cite{Javanmardi:2015sfa}.\footnote{Other indications of anisotropy from quasars and GRBs \cite{Luongo:2021nqh}, lensed quasars \cite{Krishnan:2021dyb} etc, all correlated with the CMB dipole direction, have been reviewed elsewhere \cite{Aluri:2022hzs}.}

As is evident from Figs.~\ref{fig:cumulative_qm} and~\ref{fig:scan1}, there is a systematic increase in the isotropic component of the deceleration parameter $q_\text{m}$ in analysis C1 (in which only cosmological parameters are estimated, employing distance moduli which are already incorporate light-curve standardisation), compared to analysis C2 (in which the cosmological and light-curve correction  parameters are estimated simultaneously). This is as expected since the Phillips-Tripp corrected magnitude $m_{B\text{corr}}$ is obtained allowing sample and redshift-dependence in the stretch ($x_1$) and colour ($c$) standardisation, as advocated in Ref.~\cite{Rubin:2016iqe}, which has the effect of enhancing the inferred acceleration. However, as noted earlier, this undermines the utility of SNe~Ia as standardiseable candles~\cite{Mohayaee:2021jzi}. While there may be no consensus on whether $x_1$ and $c$ should be allowed to be sample- and redshift-dependent \cite{Rubin:2016iqe,Rubin:2019ywt,Colin:2019ulu,Mohayaee:2021jzi}, the observed decay with redshift of the dipole in the deceleration parameter, as predicted in the tilted universe scenario, is independent of the specific choice of how $x_1$ and $c$ are treated (see Fig~\ref{fig:shell1}). The relative size of $\bm{H}_\text{d}$ in each shell \emph{vis a vis} the uncertainty claimed by SH0ES on $H_0$ is also largely independent of this choice (see Fig~\ref{fig:shell2}).

All these studies of anisotropy in Pantheon+ supernovae reveal a consistent pattern. When the data are analysed in the CMB frame (i.e. observables are corrected only for our inferred velocity of 368.9 km\,s$^{-1}$ w.r.t. the CRF of the assumed FLRW model), the  dipole anisotropy is compatible in size and direction with the local bulk flow~\cite{Carrick:2015xza,Watkins:2023rll,Colin:2010ds}, and that reported using X-ray clusters~\cite{Migkas:2024wts}. When heliocentric observables are employed, the direction and amplitude of the dipole anisotropy  are compatible with those found earlier~\cite{Colin:2019opb}. When a study reports no statistically significant anisotropy, or a completely different direction, it is because of using observables ($z_{\text{HD}} \equiv z $) which have been corrected for both the motion of the observer, as well as the motion of the host galaxy w.r.t. the CRF  --- thus effectively \emph{isotropising} the data (as seen in Figs.~\ref{fig:shell2} and ~\ref{fig:shell1}). However, this procedure is now questionable because of the recent finding that the dipole anisotropy in the sky distribution of radio sources and quasars does not match that expected due to our peculiar motion as deduced from the CMB dipole \cite{Secrest:2022uvx, Dam:2022wwh,Wagenveld:2023kvi}, i.e. there is \emph{no} frame in which both the CMB and distant matter are both isotropic. It can no longer be argued that the dipole in the deceleration parameter can be transformed away by boosting to this mythical frame \cite{Rubin:2019ywt}.

It is noteworthy that the anisotropy in $q_0$ is maximum in the LG frame (see Fig.\ref{fig:shell1}) which may be identified with the `tilted observer' embedded in the local bulk flow.
Also when the anisotropy in $H_0$ determined using SNe~Ia as independent distance indicators is interpreted as a bulk flow~\cite{Colin:2010ds,Feindt:2013pma}, its velocity is lower than that reported by studies utilising other tracers~\cite{Watkins:2023rll,Migkas:2024wts}. This may be because SNe~Ia compilations have evolved historically through a process of outlier rejection, which has biased the samples towards the $\Lambda$CDM expectation (see Table 1 of Ref.~\cite{Mohayaee:2021jzi}).

\section{Conclusions} \label{concl}

In this paper we have studied the anisotropy in the kinematics of the expansion of the Universe with close attention to the choice of the observer frame and the peculiar velocity corrections applied to the data, using the Pantheon+ catalogue of SNe~Ia. As seen in Fig.~\ref{fig:shell2}, there is a significant anisotropy in the Hubble expansion rate in our local neighbourhood ($z=0.023-0.15$) where SNe~Ia are calibrated by Cepheids. This anisotropy is larger than the $1.0$~km\,s$^{-1}$Mpc$^{-1}$ precision claimed in the SH0ES analysis \cite{Riess:2021jrx} carried out in the   FLRW framework. It is clear however that the local Universe cannot be thus modelled to this precision. The formulation of the `Hubble tension' does not consider systematic deviations from FLRW such as the anisotropies in the Hubble expansion that we have uncovered.

Concerning cosmic acceleration, we find results using Pantheon+ consistent with those obtained earlier using the JLA catalogue \cite{Colin:2019opb}, namely that it is mainly directed along our local bulk flow. That analysis was carried out in the heliocentric frame, but as seen in Fig.~\ref{fig:qdvsznn}, the conclusion still holds in the CMB frame and is particularly pronounced in the Local Group frame. The anisotropy is minimised if the redshifts are corrected further for the peculiar velocities of the SNe~Ia wrt the CMB frame, however there is now a question as to whether this procedure is valid, given the mismatch between the CMB frame and the frame in which distant matter is isotropically distributed, as revealed by the Ellis \& Baldwin test \cite{Secrest:2022uvx}. As seen in Figs.~\ref{fig:cumulative_qm} and \ref{fig:scan1}, the monopole component of $q_0$ remains consistent with zero, unless the light curve standardisation are allowed to be sample and redshift-dependent, while maintaining that the absolute magnitude remains invariant.

It would be helpful to do further checks with the UNION3 compilation of 2087 SNe~Ia of which 707 are at redshift $z < 0.1$ \cite{Rubin:2023ovl}, when it is made public. It uses the updated SALT3 light-curve fitter, as does the DES5Y sample of 1635 SNE~Ia in the redshift range $z=0.0596-1.12$ from the Dark Energy Survey plus 194 SNe~Ia at redshift $z < 0.1$ \cite{DES:2024hip}. However the 118 SNe~Ia common to these compilations at $z < 0.1$ have a systematic offset in magnitude of $\sim 0.05$~mag between the two catalogues \cite{Efstathiou:2024xcq}. Such systematics issues will have to be resolved in order to make progress. The Zwicky Transient Facility has already detected 3628 SNe~Ia at $z < 0.3$ \cite{Rigault:2024kzb}. Hundreds of thousands of SNe~Ia will be measured in the forthcoming Legacy Survey of Space and Time at the Rubin Observatory \cite{LSSTScience:2009jmu}. There are therefore excellent prospects for establishing at high significance whether the acceleration inferred from the supernova Hubble diagram is indeed anisotropic. This would confirm that it cannot be ascribed to $\Lambda$ but probably a general relativistic effect due to the local bulk flow. More generally, the standard assumption of isotropy and homogeneity in analysing cosmological data will be better tested with the advent of surveys covering large areas of the sky, paving the way to the construction of a more realistic cosmological model. 

\bigskip

\noindent \textbf{Code availability:} The code for reproducing the results is available at this \href{https://github.com/Shin107/Anisotropy-in-Pantheon-Plus}{Github} repository: {\tt https://github.com/Shin107/Anisotropy-in-Pantheon-Plus}. 
\bmhead{Acknowledgements}
We acknowledge use of the TIFR CCHPC cluster facility. We are grateful to Zachary Lane \& David Wiltshire for providing us the covariance matrices and helpful discussions, and to Sebastian von Hausegger for a critical reading of the manuscript and helpful feedback. 
SS was supported by the UK Science \& Technology Facilities Council.
CGT was supported by the Hellenic Foundation for Research \& Innovation (H.F.R.I.), under the ``First Call for H.F.R.I. Research Projects to support Faculty members and Researchers \& the procurement of high-cost research equipment Grant'' (Project: 789). This work made use of python packages Scipy \cite{Virtanen:2019joe}, Numpy \cite{Harris:2020xlr} and Astropy \cite{Astropy:2022ucr} for analysis, and Matplotlib \cite{Hunter:2007} and 
Scienceplots \cite{SciencePlots} for plotting.


\bibliography{Pantheon+.bib}

\begin{appendices}

\section{Quadrupole in the Hubble expansion} \label{sec:APPA}

Ref.~\cite{Heinesen:2020bej} provides a general, background-independent cosmographic luminosity distance formula upto the third order for an anisotropic universe. When dealing with null curves/rays, the general multipolar Hubble parameter is \cite{Heinesen:2020bej}.
\begin{equation}
    \mathbf{\quad\mathfrak{H}(e)}= \frac{1}{3} \theta -e^{\mu}a_{\mu} + e^{\mu} e^{\nu} \sigma_{\mu \nu}\,,  \label{A1}
\end{equation}
where $\theta$  is the volume scalar, $e^{\mu}$ is the direction of observation, $a^{\mu}$ is the four-acceleration and $\sigma_{\mu\nu}$ is the shear tensor.
If we ignore the four-acceleration, but keep the shear nonzero,\footnote{Neglecting the four-acceleration means from the theoretical point of view, that peculiar motion effects are bypassed. Keeping the shear non-zero implies that there is \emph{intrinsic} spacetime anisotropy.} the residual part of Eq.~(\ref{A1}) also follows from the timelike expression (\ref{Thetaab}), after projecting it twice along a given spatial direction. With only the quadrupolar term remaining, the generalised Hubble parameter can be expressed a:
\begin{equation}
H(e)=H_\text{m} + H_{\text{q}} .ee F(z,S_{\text{q}})
\end{equation}
Here $H_\text{m}$ is the monopole component of the Hubble parameter, $H_{\text{q}}$, the quadrupole tensor and $e$ the direction of SNe~Ia. $F(z,S_{\text{q}})$ is the assumed decay function for the quadrupole.
Similarly to Ref.~\cite{Cowell:2022ehf}, we estimate the quadrupole in the Hubble parameter, along with a dipole in the deceleration parameter, within the same redshift range of $z=0.023-0.8$. We adopt the same parameterisation as for the deceleration parameter, viz. 
$q=q_\text{m}+\bm{q}_\text{d}\cdot\ \mathbf{n}\text{e}^{-z/S_{\text{dip}}}$, with the dipole direction fixed to the CMB dipole direction. As for the Hubble parameter, we parameterise it following  Ref.~\cite{Cowell:2022ehf} as:
\begin{equation}
    H = H_\text{m} [1+(\lambda_1 \cdot \text{cos}^2 \theta_1+\lambda_2 \cdot \text{cos}^2\theta_2-(\lambda_1+\lambda_2) \cdot \text{cos}^2 \theta_3] \text{e}^{-z/S_\text{q}}
\end{equation}
with the direction of quadrupole fixed to the one obtained in Ref.~\cite{Parnovsky:2012uk}, and $S_{\text{q}}$ fixed to values as shown in Tables \ref{tab:quad_HEL_C1} to \ref{tab:quad_HD_C2}.
For consistency with Ref.~\cite{Cowell:2022ehf} we do not reverse any bias corrections in this analysis.The last row in all the tables corresponds to  the case where only a dipole in the deceleration parameter is fitted.

We find that the evidence for a quadrupolar component to the Hubble expansion is significantly weaker than that for the dipole in the deceleration parameter. Additionally, the evidence for a quadrupole is stronger, compared to the heliocentric frame, when analysed in the CMB frame with SNe~Ia peculiar velocity corrections incorporated, i.e. the CRF ($z_\text{HD}$). Simultaneously estimating the quadrupole in the Hubble parameter and the dipole in the deceleration parameter does reduce the magnitude of the dipole in the heliocentric frame, as was noted in Ref.~\cite{Cowell:2022ehf}. However, in the CMB frame with peculiar velocity corrections, the amplitude of the dipole is observed instead to increase.

\begin{table}[ht!]
\begin{tabular}{ |c|c|c|c|c|c|c|c|c|c|c|c|c|c|} 
  \hline
   $S_{\text{q}}$ & $H_\text{m}$  & $\lambda_1$&   $\lambda_2$   & $\bm{q}_\text{d}$ &$S_{\text{dip}}$ &$q_\text{m}$&$j_0-\Omega_{\text{k}}$ &  $\alpha$ (3D.O.F.)   \\
   \hline
0.03/$\ln{2}$&71.0&0.017&-0.014&-1.8&0.067&-0.39&0.53 &  0.17\\
0.06/$\ln{2}$ & 71.0 &0.016&-0.008&-1.9&0.065&-0.39&0.50 & 0.4\\
0.1/$\ln{2}$&  70.9 &0.014&-0.005&-1.9
&0.064&-0.390&0.55 & 0.59\\
- &71.1 & - &- &-1.7&0.069&-0.423&0.648&\\
\hline
\end{tabular}
\caption{Parameters describing a quadrupolar component to the Hubble expansion rate and a dipole in the deceleration parameter in the heliocentric frame from analysis C1 (wich employs sample and redshift-\emph{dependent} light curve standardisation). Each row corresponds to a different fixed $S_{\text{q}}$ value, with the final row representing the case where the quadrupole is set to zero. $\alpha$ denotes the statistical significance with which the `no-quadrupole' hypothesis is rejected according to the likelihood ratio, using Wilks' theorem (with 3 d.o.f.). (when the dipole direction is fixed).}
\label{tab:quad_HEL_C1}
\end{table}

\begin{table}[ht!]
\begin{tabular}{ |c|c|c|c|c|c|c|c|c|c|c|c|c|c|} 
  \hline
   $S_{\text{q}}$ & $H_\text{m}$  & $\lambda_1$&   $\lambda_2$   & $\bm{q}_\text{d}$ &$S_{\text{dip}}$ &$q_\text{m}$&$j_0-\Omega_{\text{k}}$ & $\alpha$ (3 D.O.F.)  \\
   \hline
0.03/$\ln{2}$&70.0 &0.021&-0.014&-4.2&0.028&-0.27&-0.02&0.21\\
0.06/$\ln{2}$ & 69.9&0.018&-0.008&-4.5&0.027&-0.26&-0.02&0.39\\
0.1/$\ln{2}$&  69.9 &0.016&-0.005&-4.6&0.026&-0.27&0.01&0.52\\
 -&70.2& -&-&-3.7&0.029&-0.30&-0.08&\\
\hline
\end{tabular}
\caption{Parameters describing a quadrupolar component to the Hubble expansion rate and a dipole in the deceleration parameter in the heliocentric frame from analysis C2 (which employs sample and redshift-\emph{independent} light curve standardisation). Each row corresponds to a different fixed $S_{\text{q}}$ value, with the final row representing the case where the quadrupole is set to zero. $\alpha$ denotes the statistical significance with which the `no-quadrupole' hypothesis is rejected according to the likelihood ratio, using Wilks' theorem (with 3 d.o.f.). (when the dipole direction is fixed).}
\label{tab:quad_HEL_C2}
\end{table}

\begin{table}[ht!]
\begin{tabular}{ |c|c|c|c|c|c|c|c|c|c|c|c|} 
  \hline
   $S_{\text{q}}$ & $H_\text{m}$  & $\lambda_1$&   $\lambda_2$   & $\bm{q}_\text{d}$ &$S_{\text{dip}}$ &$q_\text{m}$&$j_0-\Omega_{\text{k}}$  & $\alpha$ (3 D.O.F.) \\
   \hline
.03/$\ln{2}$&71.6&0.031&-0.008&1.9&0.0239&-0.48&0.92&0.82\\
0.06/$\ln{2}$ & 71.5&0.023&-0.003&  1.9&0.0239&-0.48&0.92&1.11\\
0.1/$\ln{2}$&  71.5&0.019&-0.0006&1.9&0.0239&-0.49&0.99&1.66\\
 - &71.9& - &- &2.4&0.0239&-0.526&1.12&\\
\hline
\end{tabular}
\caption{Parameters describing a quadrupolar component to the Hubble expansion rate and a dipole in the deceleration parameter in the CRF (CMB frame with peculiar velocity corrections) from analysis C1 (wich employs sample and redshift-\emph{dependent} light curve standardisation). Each row corresponds to a different fixed $S_{\text{q}}$ value, with the final row representing the case where the quadrupole is set to zero. $\alpha$ denotes the statistical significance with which the `no-quadrupole' hypothesis is rejected according to the likelihood ratio, using Wilks' theorem (with 3 d.o.f.). (when the dipole direction is fixed).}
\label{tab:quad_HD_C1}
\end{table}

\begin{table}[ht!]
\begin{tabular}{ |c|c|c|c|c|c|c|c|c|c|c|c|c|c|} 
  \hline
$S_{\text{q}}$ & $H_\text{m}$  & $\lambda_1$&   $\lambda_2$   & $\bm{q}_\text{d}$ &$S_{\text{dip}}$ &$q_\text{m}$&$j_0-\Omega_{\text{k}}$ & $\alpha$ (3D.O.F.)   \\
\hline
0.03/$\ln{2}$&70.5&0.034&-0.009&0.8&0.07&-0.367&0.39&0.91\\
 0.06/$\ln{2}$ & 70.5 &0.024&-0.003&0.8&0.07&-0.368&0.412&1.09\\
0.1/$\ln{2}$&  70.5&0.019&0.0003&0.8&0.07&-0.38&0.478&1.19\\
- &70.8 & - &- &1.0&0.06&-0.414&0.589&\\
\hline
\end{tabular}
\caption{Parameters describing a quadrupolar component to the Hubble expansion rate and a dipole in the deceleration parameter in the CRF (CMB frame with peculiar velocity corrections) from analysis C2 (wich employs sample and redshift-\emph{independent} light curve standardisation). Each row corresponds to a different fixed $S_{\text{q}}$ value, with the final row representing the case where the quadrupole is set to zero. $\alpha$ denotes the statistical significance with which the `no-quadrupole' hypothesis is rejected according to the likelihood ratio, using Wilks' theorem (with 3 d.o.f.). (when the dipole direction is fixed)..}
\label{tab:quad_HD_C2}
\end{table}

\section{Treatment of covariance matrices} \label{sec:AMMA}
The Pantheon+ covariance matrix \cite{Brout:2022vxf} accounts for both statistical and systematic uncertainties, however, unlike for the Joint Lightcurve Analysis (JLA), it does not allow for decomposition into individual systematic components, as their individual covariance matrices are not explicitly provided. Ref.~\cite{Lane:2023ndt} therefore augmented the  Pantheon+ covariance matrix with additional components thus:
\begin{equation}C=C_{\text{fit}}+C_{\text{stat}}+C_{\text{dupl}}+C_{\text{FITOPTS}}+C_{\text{MUOPTS}},
\end{equation}
where:
\begin{trivlist}

\item
$C_{\text{fit}}$: The SALT2 fit covariance matrix, derived from the terms in the Pantheon+ table.

\item
$C_{\text{stat}}$: The statistical covariance matrix, which incorporates the terms $\sigma^2_\text{lens}$ and $\sigma^2_{z}$. It is important to note that, unlike Pantheon+, this matrix does not include $\sigma^2_\text{{floor}}$, $\sigma^2_\text{scat}$, $\sigma^2_\text{gray}$ and $\sigma^2_\text{vpec}$, as these terms depend on the assumed cosmological model.

\item
 $C_\text{dupl}$: The contribution from duplicate observations.
 
\item
 $C_\text{FITOPTS}$ and $C_\text{MUOPTS}$: These correspond to systematics and are calculated as  $C^{ij}_\text{syst}=\sum_{\psi} \sigma_{\psi}\partial_{\psi}\mu^{i}\partial_{\psi}\mu^{j}$ where the sum is over all systematic uncertainties $\psi$. This formula accounts for the impact of various calibration uncertainties of the light curve parameters over different systematic effects (excluding the calibration associated with peculiar velocities).
 
 \end{trivlist}
 
\begin{figure}[htb]
    \centering
    \includegraphics[scale=0.5]{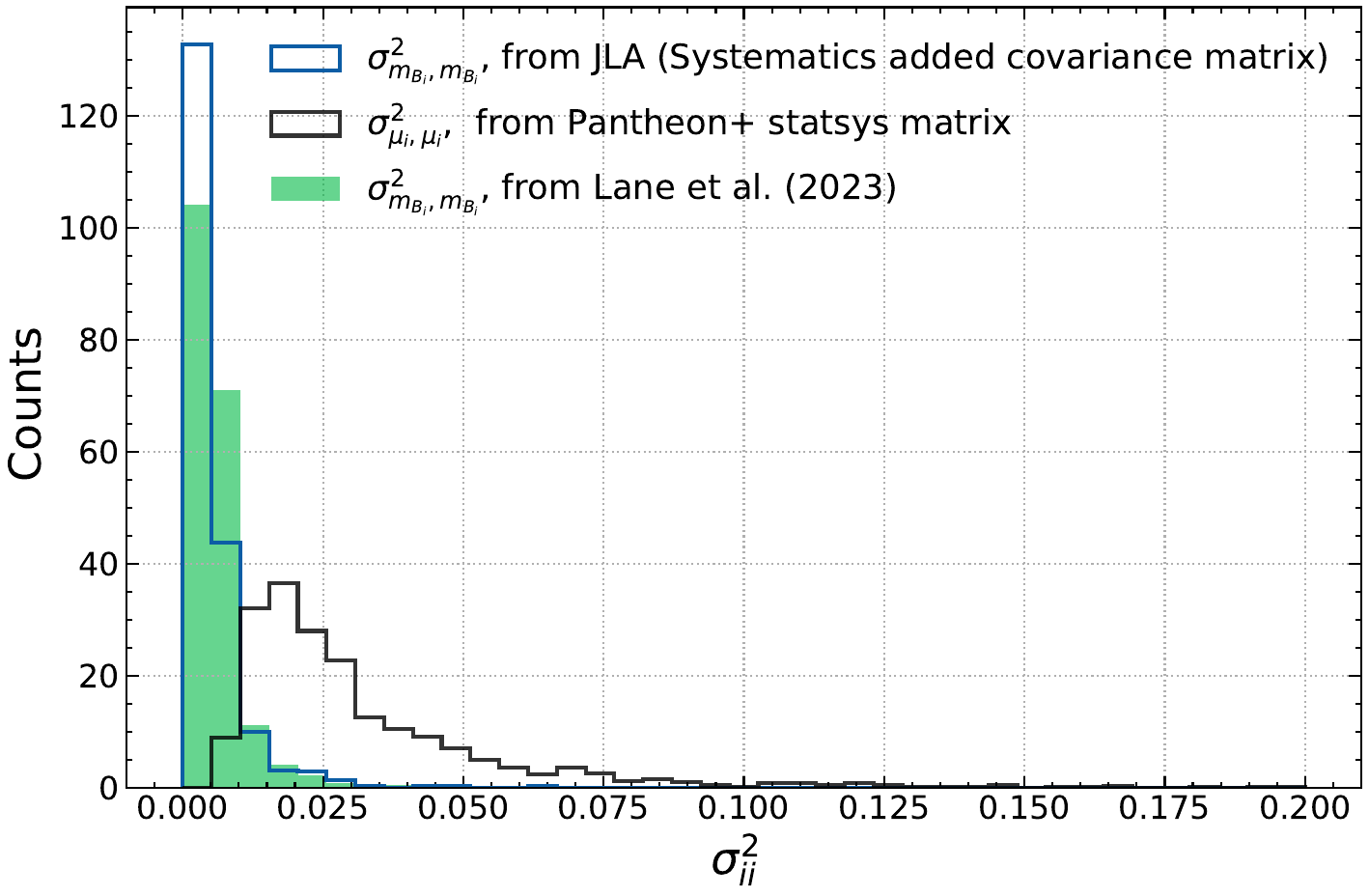}
    \caption{Histogram comparing diagonal components of the statistical+systematic covariance matrix employed in JLA \cite{SDSS:2014iwm} and in Pantheon+ \cite{Brout:2022vxf} with that constructed in Ref.~\cite{Lane:2023ndt} for the Pantheon+ compilation.}
    \label{fig:histcomp}
\end{figure}

 Fig. \ref{fig:histcomp} compares the diagonal $\sigma_{m_{B_i},m_{B_i}}$ component of JLA \cite{SDSS:2014iwm} with the corresponding $\sigma_{\mu_{i},\mu_{i}}$ component of Pantheon+ statistical+systematic covariance \cite{Brout:2022vxf}, and the matrix constructed in Ref.~\cite{Lane:2023ndt}. 

-------------------------------------------------------------------------

\end{appendices}

\end{document}